\documentclass[
superscriptaddress,
preprint,
amsmath,amssymb,
aps,
prfluids
]{revtex4-2}
\usepackage{color}
\usepackage[colorlinks=true,citecolor=blue,urlcolor=black]{hyperref}
\usepackage{graphicx}
\usepackage{epstopdf}
\usepackage{dcolumn}
\usepackage{bm}
\usepackage[english]{babel}

\begin{document}


\title{A parametric study of the broadband shock-associated noise in
supersonic jets via semi-empirical modeling}

\author{Binhong Li}
\author{Benshuai Lyu}%
 \email{b.lyu@pku.edu.cn}
\affiliation{%
State Key Laboratory of Turbulence and Complex
    Systems, School of Mechanics and Engineering Science, Peking University, 
    5 Yiheyuan Road, Haidian District, Beijing 100871, China\\
}%



\begin{abstract}
A semi-empirical model is developed in this paper to predict the broadband
shock-associated noise (BBSAN) generated by shock-instability interaction
(SII) in imperfectly expanded supersonic jets. The model
makes use of a semi-empirically modified Pack's model that accounts for the decay in both shock
amplitude and shock spacing and a Gaussian wave-packet model for the instability
waves. The near-field pressure perturbation due to the SII is treated as a
boundary value for the Helmholtz equation, which is subsequently solved to
predict the far-field acoustic spectra and directivity patterns. A comprehensive
parametric study is conducted to reveal the effects of the key parameters on the
acoustic spectral and directivity features. 
It is found that
decreasing the instability-wave decay rate narrows the spectral bandwidth and
the major lobes in directivity patterns, 
 while variations in shock spacing shift the spectral peak frequency and the major radiation angle.  Mechanisms of such changes are discussed based on the model. {Further validation against multiple experimental datasets demonstrates that incorporating more realistic parameters in the model—particularly those accounting for the shock spacing and amplitude decays— considerably improves its prediction accuracy and physical consistency. The improved model successfully reproduces several key spectral features observed in experiments; these include, for example, the peak frequency and the tendency of bandwidth contradiction as the observer angle increases. Moreover, the predicted directivity patterns closely match the experiments outside the shallow-angle region dominated by jet mixing noise. In particular, it captures the major radiation lobes and their frequency-dependent amplitude and shape variations.}  \end{abstract}

\maketitle
\newpage

\section{Introduction}
\label{sec:intro}
High-speed aircraft have the potential to reduce flight times significantly, representing a key technological arena in the aviation industry~\citep{highAircraft}. Such aircraft are often powered by supersonic jet engines, and the supersonic jet exhaust from its engine may operate under off-design conditions {\citep{broad_band_2013}}. In such conditions, the broadband shock-associated noise (BBSAN) can be generated due to the interaction between shock waves and shear-layer instability waves (SII) \citep{1995_annu_tam}. When the velocity of the aircraft further increases,  the BBSAN becomes increasingly dominant, causing problems such as structural fatigue or hearing loss for both passengers on board and individuals on the ground {\citep{Miller_2016}}.

To investigate the generation mechanisms of the BBSAN and develop effective noise
control strategies, extensive studies were conducted using experimental
measurements \citep{1982NormSeiner,TDNorum_AIAA, farfield_rectangular,
1986Seiner, E.GUTMARK1990} and numerical simulations \citep{broad_band_2013,
2019Wong, Xiangru_2019, absolute_instability}. These studies showed that BBSAN
mainly occurs in a relatively low-frequency regime and primarily radiates
upstream of the jet. Its main sources appeared to be located at several jet
diameters downstream of the nozzle, where instability waves reach high
intensity.

Compared to experiments and simulations, analytical modeling not only requires 
significantly lower computational resources, but also provides an essential
method for examining the underlying physical mechanisms of the BBSAN. 
To develop such analytical models, it is essential to first develop accurate models for both shock and instability waves. 

Regarding shock structures, studies showed that when the jet operated under slightly off-design conditions~\citep{1982Tam}, Pack's model \citep{1950Pack}, based on the vortex sheet assumption, reliably predicted shock spacing \citep{X.D.Li} and shock-induced velocity distributions within a single shock structure \citep{1982NormSeiner}.
Following Pack’s work, similar models for predicting shock spacing in non-axisymmetric and beveled jets were developed by \citet{1988Tam_shockspacing} and \citet{1996Tam_JOA}, respectively. Instead of using the vortex sheet assumption, \citet{1985Tam} introduced a multiple-scale model that considered a slowly diverging jet. This model successfully captured the fine structure of shock cells and the evolution of shock intensity along the streamwise direction. More recently, \citet{2024Song_nonlinear} developed a nonlinear model to examine the effects of nonlinearity on shock structures. They found that while nonlinearity had a minor influence on shock intensity, it did not affect shock spacing.
{Despite these advancements, Pack’s model remains widely used due to its simplicity and relatively high accuracy. It performs well when calculating shock spacing in slightly off-design jets and representing shock-induced perturbations within a single shock structure, although it cannot resolve the fine structure of shock cells nor the downstream decrease in shock intensity and spacing well-known in experiments \citep{1982NormSeiner}.

Regarding shear-layer instability waves, models based on the vortex-sheet \citep{196batchelor} or the parallel-flow assumptions~\citep{1982Alfons,
LSA_morris} appeared to capture their behavior near the jet nozzle. When the
slow divergence of the jet flow needs to be taken into account, the parabolized
stability equation (PSE) \citep{PSE_piot, Gudmundsson_jfm, nogueira2022wave} and
the WKB method \citep{1976Crighton_diverging} may be used to model the
evolution of the instability waves along the streamwise direction. These
approaches primarily focus on the linear growth stage of instability waves;
however, nonlinear effects may become significant further downstream. Under such
conditions, instability waves can be analyzed using methods such as the
nonlinear parabolized stability equation (NPSE) \citep{Yen_aiaaj}, {modified
one-way Navier–Stokes equations (OWNSE) with nonlinear forcing} 
\citep{OWNSE_2022}, and
other nonlinear theories \citep{Wu_nonlinearMachWave, Wu_annual}.  It is widely
believed that the large-scale shear-layer instability waves exhibit a
characteristic structure of wave packets due to linear and nonlinear
saturations~\citep{2013_annual_rev, 2020_jfm_wavepackets}. The amplitude of
these wave packets may be approximated by a Gaussian envelope
\citep{exponetially_Gaussian}. {Note that such wave packets may occupy a
large spatial region, including the area where the BBSAN primarily occurs.}

The BBSAN may then be modeled by using appropriate models of the shock and
instability waves. Such models may be used to {address the two key questions, i.e. (1) characterizing the SII within the jet plume and (2)
predicting the acoustic waves generated by the SII in both the near and far
fields.} For example, following the pioneering work of \citet{1973Harper},
\citet{1982Tam} developed a semi-analytical model that describes the SII {in
the jet plume}  
by multiplying the
perturbations induced by shock and instability waves. In this model, the shock
structure was obtained using Pack’s model, while the instability waves were
represented as a linear superposition of normal modes with random amplitude
functions. The study revealed that some disturbance components attain supersonic
phase speeds along the jet due to the SII, leading to Mach wave radiation. Both
the radiation angle and the frequency of the BBSAN were calculated using the
Mach angle relation.

Building on this, \citet{1988Tam} proposed a {semi-empirical} 
model to predict both
near-field and far-field acoustic emissions due to the SII. This model assumed that
the acoustic wave followed a similar form to the SII described in
\citet{1982Tam}, with parameters such as the convection velocity of the
instability wave and the wave-packet half-width determined through linear
stability analysis or experimental measurements. The predicted results showed
good agreement with experimental data.

\citet{2005Lele} further developed a phased-array theory, {by considering phase arrays of localized and distributed sound sources.}   
Similar to
\citet{1988Tam}, this model treated the instability wave as a wave packet and
modeled the shock cell structure using Pack’s model. However, it determined the
SII in the jet plume by reformulating the
Euler equations into a Helmholtz equation with a source term. This source term,
resembling the quadrupole form proposed by \citet{1952lighthill}, was determined 
by the perturbations induced by shock and instability waves. The far-field
acoustic wave was subsequently obtained by {convoluting}  
the source term with the free-space Green's function. \citet{2019Wong} further investigated
the effects of coherence decay of the wave-packet model on the BBSAN. Results
showed that this decay mainly changed the acoustic spectra at relatively high
frequencies.

{Previous studies have shown that the wave-packet models can capture the
essential behavior of the instability waves, while Pack's model provides
a useful first-order approximation of shock structures. However, Pack's
model neglects important downstream evolution—particularly the gradual decay of
both shock amplitude and spacing—which may lead to pronounced discrepancies in
regions far from the nozzle. As a result, the effects of these parameters on the
BBSAN are yet to be understood.}  To bridge the gap, we model the shock
structures using a semi-empirically modified Pack model that includes the
variation of shocking spacing and amplitudes along the streamwise direction and
propose a semi-empirical model of the BBSAN that aims to model both the shock
and instability waves as realistically as possible. Following the approach of
\citet{1988Tam}, the SII in the jet plume is modeled as the product of
disturbances induced by shock and instability waves. However, the far-field
acoustic wave is obtained by solving the Helmholtz equation with an assumed
boundary value from the SII, rather than directly assuming the same form of the
SII.

The structure of this paper is as follows: Sec. \ref{subsec:model_derivation} presents a detailed
derivation of the semi-empirical model, while Sec. \ref{subsection:2.2} discusses several key physical
parameters, {including the decay rates of the shock and instability amplitudes.} 
 In Sec. \ref{sec:3}, the predicted directivity
patterns and frequency spectra of the BBSAN are shown. In addition, a parametric study is conducted to study and discuss the effects of the non-dimensional parameters on the BBSAN. A comparison with experimental data is then shown in Sec. \ref{sec:3.3}. Finally, conclusions are drawn in Sec. \ref{sec:4}. 


\section{Model formulation} \label{sec:types_paper}
\begin{figure}
   \centering
   \includegraphics[width = 0.65\textwidth]{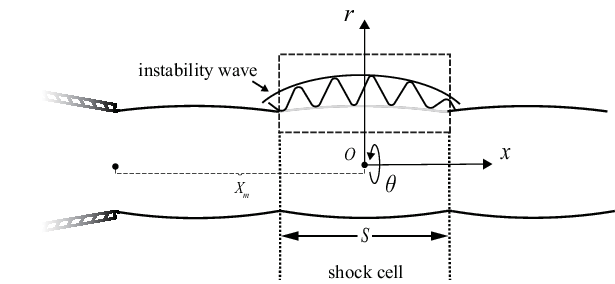}
   \caption{Schematic of the jet flow in a cylindrical
   coordinate frame. The origin is fixed on the jet center line, while $x $, $r $, and $\theta$ represent the streamwise, radial, and azimuthal coordinates, respectively.  Note that the 
instability wave reaches its maximum intensity at $x=0$ while the nozzle is located at $x=-X_m$.} 
\label{fig:examplee1}
\end{figure}

\subsection{The parametric model}
\label{subsec:model_derivation}
As illustrated in Fig. \ref{fig:examplee1}, 
the jet is issued from a circular
nozzle and continues to develop along the streamwise
direction. 
 The jet velocity at the nozzle exit is denoted by $\tilde{U}_e$,
while the velocity of the fully expanded jet flow is represented as
$\tilde{U}_j$. The diameter of the fully expanded jet flow, $\tilde{D}_j$, may
be larger or smaller than the nozzle diameter $\tilde{D}$, depending on whether the flow is
under-expanded or over-expanded.

The mean density and static temperature inside the jet flow are denoted by
$\tilde{\rho}_{0-}$ and $\tilde{T}_{0-}$, respectively, while the corresponding
parameters outside the jet are represented by $\tilde{\rho}_{0+}$ and
$\tilde{T}_{0+}$. The temperature ratio between $\tilde{T}_{0-}/\tilde{T}_{0+}$ is defined as $\nu$. The speeds of sound inside and outside the jet can be then
calculated by $\tilde{a}_{0\pm}=\sqrt{\gamma \tilde{p}_0/\tilde{\rho}_{0\pm}}$
when a perfect gas is assumed. Here $\gamma$ denotes the specific heat ratio and
$\tilde{p}_0$ represents the mean pressure, which is assumed to be the same
inside and outside the jet. Two Mach numbers are defined for the jet flow, i.e.
$M_{-} = \tilde{U}_j / \tilde{a}_{0-}$ and $M_{+} = \tilde{U}_j /
\tilde{a}_{0+}$. Both represent the jet Mach number but are calculated based on
the speed of sound in different regions. In what follows, we non-dimensionize
relevant variables using $\tilde{D}_j$, $\tilde{U}_j$, $\tilde{\rho}_{0-}$ and
$\tilde{T}_{0-}$. {We use the symbols with a tilde to represent 
dimensional variables, while those without to denote non-dimensional variables.}

To simplify the formulation, we construct a cylindrical coordinate centered at a
point on the jet center line. The non-dimensional streamwise, radial, and
azimuthal coordinates are denoted as $x$, $r$, and $\theta$, respectively. As
illustrated in Fig. \ref{fig:examplee1}, {the instability wave attains its
maximum intensity at $x=0$.} {The distance from the nozzle center to the origin
of the coordinate is defined as $X_m$. Given that the shock-induced pressure
amplitude typically peaks immediately downstream of the nozzle,  $X_m$ can also
be interpreted as the spatial offset between the peak locations of the shock and
instability waves amplitudes.}

{A modified Pack's model} is used to describe the shock structures near the lip line, i.e., a modified cosine function
representing the shock-induced pressure {variation}  
$p_s$ along the streamwise
direction. Note that the {variation} 
$p_s$ reaches its maximum at the nozzle exit.  
Therefore, we have
\begin{equation}
p_s=\mathcal{A}_s \cos[2\pi (x+X_m)/S],
\label{equ:ps}
\end{equation}
where the coefficient $\mathcal{A}_s$ denotes the shock amplitude, which
generally decreases downstream of the nozzle, as observed by
\citet{1982NormSeiner}. Such a decay will be discussed in detail in Fig. \ref{fig:exampl3}.  
As a starting point, we assume a linear decrease in $\mathcal{A}_s$
downstream of the jet flow:
\begin{equation}
\mathcal{A}_s=\mathcal{A}_{s0}(1- \epsilon_s (x+X_m)),
\end{equation}
where $\epsilon_s$ represents the shock amplitude decay rate.  
The parameter $S$ denotes the shock spacing, with the initial spacing immediately downstream of the nozzle exit  $S_0$ determined via Pack's model. Experimental studies have also
shown that $S$ decreases gradually along the jet axis \citep{1986Tam_Proposed}. Similar to
$\mathcal{A}_s $, we assume a linear decrease in  $S$, with the decay rate given
by $\sigma$, 
\begin{equation}
S=S_0(1-\sigma (x+X_m))=\frac{\pi}{2.4048}\sqrt{M_{-}^2-1}(1-\sigma (x+X_m)).
\end{equation}
Discussions on the use of linear decays will be shown in Sec.~\ref{subsection:2.2}.

To describe the wave packet of instability waves around the nozzle lip line, we use a Gaussian function \citep{2013_annual_rev, 2019wavepacket}, 
\begin{equation}
p_i=\mathcal{A}_i \mathrm{e}^{-(\epsilon_i x)^2}\mathrm{e}^{\mathrm{i}\alpha x-\mathrm{i}\omega t+\mathrm{i}n\theta},
\label{equ:pi}
\end{equation}
where $\mathcal{A}_i$ denotes the instability amplitude, $\epsilon_i$ the decay
rate, $\omega$ the angular frequency, $n$ the azimuthal mode, and $\alpha$ represents the streamwise
wavenumber of instability waves.  Here, $\alpha$ is a real number that governs
the propagation of the instability wave along the streamwise direction. For
brevity, the time-harmonic term $\mathrm{e}^{-\mathrm{i}\omega t}$ and {the azimuthal-dependence term $\mathrm{e}^{\mathrm{i}n \theta}$ } are omitted
in the following discussions.

Similar to \citet{1988Tam}, we start by modeling SII as a simple product of $p_s$ and $p_i$, also consistent with earlier analytical models {\citep{MyOwn_2}}, i.e.
\begin{equation}
\mathcal{A}_s\mathcal{A}_i  \cos(2\pi (x+X_m)/S)\mathrm{e}^{-(\epsilon_i x)^2}\mathrm{e}^{\mathrm{i}\alpha x}.
\label{equ:2.5}
\end{equation}
Acoustic waves are generated due to the SII. Its induced pressure {variation} outside the jet, $p_a$, may be solved using the Helmholtz equation, i.e
  \begin{equation}
     ( \mathbf{\nabla}^{2}+\omega^2M_{+}^2)p_a=0.
      \label{equ_sound_homogenous}
   \end{equation} 
{The perturbations induced by the SII around the jet lip line, i.e., Eq. (\ref{equ:2.5}), are regarded as the boundary value of the resulting acoustic wave. 
Within the nozzle region, i.e., $x<-X_m$, the boundary value is set to be 0 considering that the instability waves are very weak. Furthermore, when the amplitude of the shock structures decays to 0, specifically, for $x>-X_m+1/\epsilon_s$ or $x>-X_m+1/\sigma$, the source term is also set to be 0. {It should be noted that by doing so, the scattering effects from the nozzle lip are not taken into account and are therefore excluded from the present model formulation.} Given that $\epsilon_s$ is typically larger than $\sigma$ (as can be seen in Figs. \ref{fig:exampl3} and \ref{fig:exampl4}), we have} 
  \begin{equation}
 p_a(x,1/2)=\left\{
\begin{array}{rcl}
 &0,    &{x\leq -X_m,} \\
  &\mathcal{A}_s\mathcal{A}_i \mathrm{e}^{-(\epsilon_i x)^2+\mathrm{i}\alpha x}\cos{2\pi (x+X_m)}/{S},   &{-X_m<x<-X_m+1/\epsilon_s,}\\
   &0,    &{x\geq -X_m+1/\epsilon_s.} \\
 \end{array}
 \right.
 \label{equ:boundary_value}
   \end{equation}
    
Applying the Fourier transform to Eq. (\ref{equ_sound_homogenous}) along the
streamwise and azimuthal direction, and then performing the inverse Fourier
transform, we can calculate $p_a$ as
          \begin{equation}
       p_a(x,r)=\frac{1}{2\pi}\int_{-\infty}^{+\infty}F(k)\frac{H^{(1)}_n(\gamma_o r)}{H^{(1)}_n(\gamma_o /2)}\mathrm{e}^{-\mathrm{i}k x}\mathrm{d}k, 
        \label{equ: numerical integration}
\end{equation}
where $k$ is the streamwise wavenumber, {$\gamma_o=\sqrt{\omega^2 M_{+}^2-k^2}$}, {$H^{(1)}_n$} the $n$th-order Hankel
function of the first kind, and $F(k)$
denotes {an amplitude function related to $k$.}
The function $F(k)$ may be obtained from the SII on jet lip line via
\begin{equation}
F(k)=\int_{-\infty}^{\infty} p_a(x,1/2)\mathrm{e}^{\mathrm{i}kx} \mathrm{d}x.
\label{equ:Fk}
\end{equation}

From Eq. (\ref{equ:boundary_value}), if the shock structures are considered strictly periodic along the streamwise direction, i.e. $\sigma=\epsilon_s=0$, we can evaluate the integration in Eq. (\ref{equ:Fk}) analytically and write $F(k)$ as 
\begin{equation}
\begin{aligned}
F(k)
=-\frac{\sqrt{\pi}\mathrm{i}}{2\epsilon_i}\mathcal{H} &\mathcal{A}_{s0}\mathcal{A}_i \mathrm{e}^{-\left[X^2+(2 \pi)^2\right] / 4 \epsilon_i^2 S_0^2} \times \\ 
&\left\{ \mathrm{exp} \left[ -\pi X/\epsilon_i^2S_0^2+2\pi\mathrm{i}X_m/S_0\right] \mathrm{erfi}\left( \frac{(X+2\pi+2\mathrm{i}\epsilon_i^2S_0^2 x/S_0}{2\epsilon_i S_0}\right) \right.\\
&\quad \quad +\left.\mathrm{exp}\left[ \pi X/\epsilon_i^2S_0^2-2\pi\mathrm{i}X_m/S_0\right] \mathrm{erfi}\left( \frac{X-2\pi+2\mathrm{i}\epsilon_i^2S_0^2 x/S_0}{2\epsilon_i S_0}\right)\right\}\bigg|_{-X_m}^{\infty},
\end{aligned}
\label{equ:Fk_degenrated}
\end{equation}
where $\mathrm{erfi}$ represents the imaginary error function and the parameter $X$ is defined by $X=(\alpha+k)S_0$

 {By approximating the Hankel function $H^{(1)}_n(\gamma_o r)$ as  $\sqrt{\frac{2}{\pi\gamma_{o}r}}\mathrm{e}^{\mathrm{i}(\gamma_{o}r-n\pi/2-\pi/4)}$ in the far-field and subsequently using the saddle point method to estimate the integration in Eq. (\ref{equ: numerical integration})} \citep{Noble}, we can write the acoustic wave radiating to $(R,\psi)$ as
\begin{equation}  
p_a(R,\psi)\sim  F(k_0) \frac{1}{H^{(1)}_n(\omega M_+\sin\psi /2)} \frac{\mathrm{exp}\left({\mathrm{i}\omega M_{+}R}\right)}{R}, 
\label{equ:result}
\end{equation}
{where $k_0=-\omega M_+ \cos\psi$ represents the saddle point}, and $R$ and $\psi$ denote the distance from the source and the observer angle relative to the downstream direction, respectively. {Further details on the saddle point method can be found in \citet{ModernMethods}}. The function $F(k_0)$ can be readily calculated using Eqs. (\ref{equ:boundary_value}) and (\ref{equ:Fk}), and if $\sigma=\epsilon_s=0$, $F(k_0)$ can be directly obtained via Eq. (\ref{equ:Fk_degenrated}).

\subsection{Determination of model parameters}
\label{subsection:2.2}

Before Eq. (\ref{equ:result}) can be used to calculate the far-field sound, the
coefficients in Eq. {(\ref{equ:boundary_value})}
, e.g. $\mathcal{A}_{s0}$ and $\mathcal{A}_i$, 
need to be determined. Regarding
$\mathcal{A}_{s0}$, i.e., the intensity of the first shock structure, it is primarily
influenced by the {nozzle pressure ratio} of the jet.
Following the approach proposed by Tam, the shock intensity is evaluated using the quantity $|M_{-}^2-M_d^2|$, where $M_d$ denotes
the designed Mach number of the nozzle. This approach is also used in the present
study, i.e. 
\begin{equation}
\mathcal{A}_{s0}=|M_{-}^2-M_d^2|^2.
\end{equation}

\begin{table}
    \begin{center}
    \begin{tabular}{lc}
    \toprule
        Parameters   & Physcial meaning\\[3pt]
        $\epsilon_i S_0$   &Exponential amplitude decay rate of instability wave {per shock spacing}\\
       $\epsilon_s S_0$ &  Amplitude decay of shock wave {per shock spacing}\\
         $\sigma S_0$ &  Spacing decay of shock wave {per shock spacing}\\
         $X_m/S_0$ &  Spatial mismatch of instability and shock waves\\
    \end{tabular}
    \caption{{The physical meanings underlying non-dimensional parameters.}}
    \label{tab:1}
    \end{center}
\end{table}

The amplitude of the instability wave $\mathcal{A}_i$, on the other hand, is likely to depend on the parameters such as $\omega$, $M_{-}$, and $\nu$. {As shown by }\citet{Gudmundsson_jfm}, the experimentally  
measured wavelength and amplitude envelope of the wavepackets structure of the near-field instability waves agree well with the predictions from the PSE. Therefore, to evaluate $\mathcal{A}_i(\omega, M_{-}, \nu)$, we assume such a conclusion is also valid in the supersonic jet and perform a linear stability analysis using PSE. To determine the initial amplitude of the PSE solution at $x=-X_m$, we further assume a white-noise forcing amplitude at the nozzle lip~\citep{1988Tam}. The jet mean flow used to initiate the PSE calculation can be fitted from experimental data. In the absence of experimental mean flow data, the jet potential core length is estimated using the empirical relation proposed by \citet{zaman1998asymptotic},
\begin{equation}
 L_p = 7 + 0.8M_{-}^2. 
 \label{equ:empirical_formula}
 \end{equation}
The full mean flow profile is then reconstructed following the procedure outlined by \citet{LSA_morris}. Further details on the PSE methodology can be found in \citet{PSE_piot}.

\begin{figure}
   \centering
   \includegraphics[width = 0.9\textwidth]{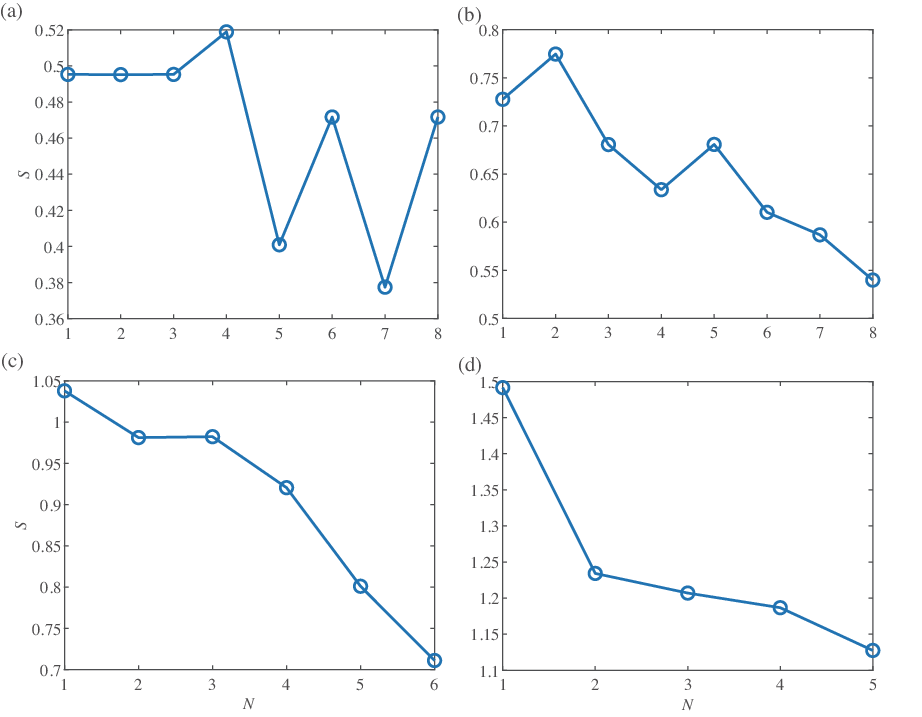}
   \caption{The obtained shock spacing from \citet{1982NormSeiner}. The designed Mach number of the nozzle is $M_d=1$ and the Mach number of the fully expanded jet is calculated via $M_{-}=\sqrt{\beta^2+1}$. (a) $\beta=0.4$; (b) $\beta=0.6$; (c) $\beta=0.8$; (d) $\beta=1$.}
\label{fig:exampl2}
\end{figure}

\begin{figure}
   \centering
   \includegraphics[width = 0.9\textwidth]{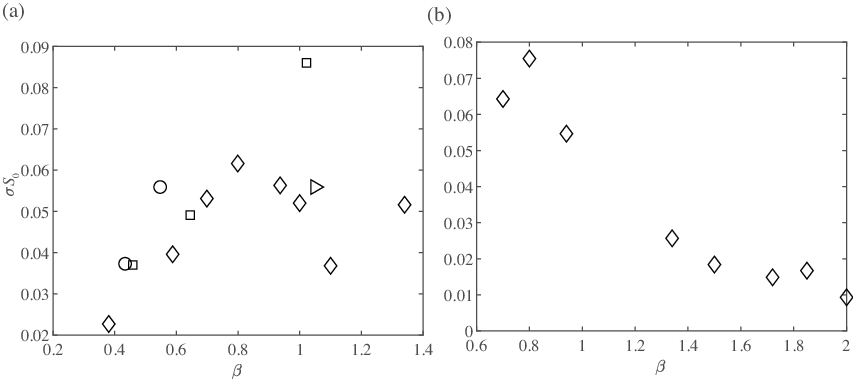}
   \caption{The decay rate of the shock spacing. $\Diamond$:
   \citet{1982NormSeiner}; $\Box$: \citet{shock_spacing_panda};
$\vartriangleright$: \citet{2014edgington}; $\circ$:
\citet{edgington_shockleakage}. (a) $M_d=1$; (b) $M_d=1.5$.}
\label{fig:exampl3}
\end{figure}

\begin{figure}
   \centering
   \includegraphics[width = 0.9\textwidth]{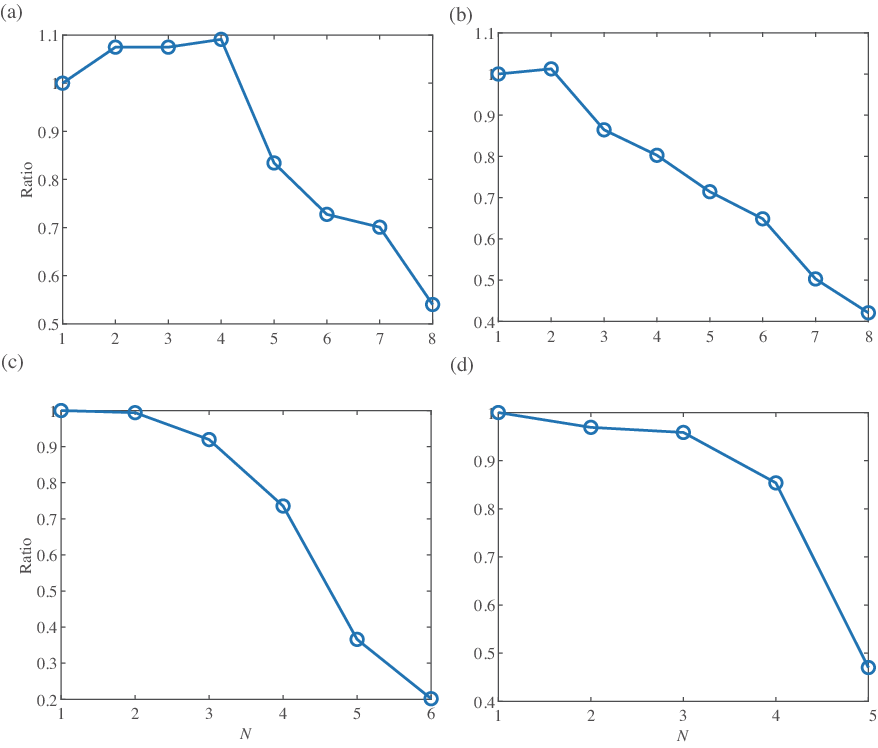}
   \caption{{The obtained shock intensities from \citet{1982NormSeiner}, which are normalized by the intensity of the first shock cell. The designed Mach number of the nozzle is $M_d=1$ and the Mach number of the fully expanded jet is calculated via $M_{-}=\sqrt{\beta^2+1}$. (a) $\beta=0.4$; (b) $\beta=0.6$; (c) $\beta=0.8$; (d) $\beta=1$.}}
\label{fig:exampl2_amp}
\end{figure}

\begin{figure}
   \centering
   \includegraphics[width = 0.9\textwidth]{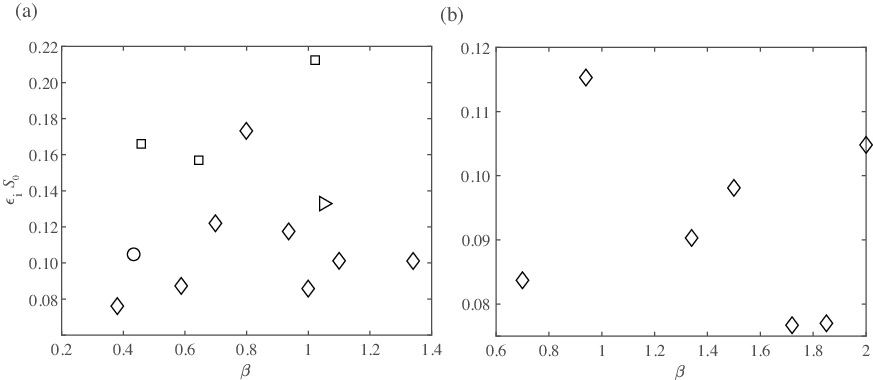}
   \caption{The decay rate of the shock amplitude. $\Diamond$: \citet{1982NormSeiner}; $\Box$: \citet{shock_spacing_panda}; $\vartriangleright$: \citet{2014edgington}; $\circ$: \citet{edgington_shockleakage}. (a) $M_d=1$; (b) $M_d=1.5$.}
\label{fig:exampl4}
\end{figure}
In addition to these two amplitude coefficients, four non-dimensional parameters, namely
\( \epsilon_i S_0 \), \( \epsilon_s S_0 \), \( \sigma S_0 \), and \( X_m / S_0 \),
govern the far-field acoustic emission. The physical meanings of these parameters are summarized in Table
\ref{tab:1}. The two coefficients related to the shock structures can be
determined through experiments. Specifically, the shock spacing and intensity can
be easily obtained from pressure or streamwise velocity measurements along the
jet's center or lip line. Typical experimental results for shock spacing are
shown in Fig. \ref{fig:exampl2}, where \( N \) represents the number of
shocks along the streamwise direction {starting from the nozzle exit} 
 and
$\beta$ is defined as $\sqrt{M_{-}^2 - 1} $.  
 It is observed that the
shock spacing generally decreases along the streamwise direction. The spacing decay per shock structure, i.e. \( \sigma S_0 \), is computed by performing a linear fit. {Note that the linear decay trend is not evident in Fig. \ref{fig:exampl2}(a), possibly because the low nozzle pressure ratio results in a less distinct shock structure.} {Neverthless, it seems reasonable to assume a linear decay profile in other cases, at least as a starting point.}

{Figure \ref{fig:exampl3}} shows the spacing decay per unit shock obtained from
several experiments as a function of $\beta$. The designed Mach numbers \( M_d \) in Fig. \ref{fig:exampl3} are (a) \( M_d = 1 \) and (b) \( M_d = 1.5 \), respectively.
It can be observed that as the jet Mach number increases, the trend of the per-shock spacing decay \( \sigma S_0 \) varies between the two cases. For \( M_d = 1 \), \( \sigma S_0
\) generally increases with \( \beta \) when \( \beta < 0.8 \), while no
distinct trend is observed when \( \beta > 0.8 \). In contrast, for \( M_d = 1.5
\), \( \sigma S_0 \) typically decreases as \( \beta \) increases. For choked
nozzles commonly used in experiments, the per-shock spacing decay appears close to 
5\% in most cases and does not exceed 10\%. {Since no clear trend is observed for $\sigma S_0$ with respect to $\beta$, we do not wish to impose a fixed value but choose to conduct a parametric study in the following sections to evaluate its effects on the BBSAN.}  

{Similarly, the variation of shock structure intensity along the streamwise direction is presented in Fig. \ref{fig:exampl2_amp}, where a general decreasing trend is also observed.} The amplitude of the non-dimensional linear decay rate can be then fitted numerically, and the results are shown
in Fig. \ref{fig:exampl4}.  
It can be observed that as the jet Mach number varies, \(
\epsilon_s S_0 \) can reach up to 22\%, while in some cases, it can be as low as
8\%.  

{Similar to the amplitude of the instability waves, $\mathcal{A}_i$, the parameters $\epsilon_i S_0$ and $X_m/S_0$ related to the instability are obtained via the PSE method. Specifically, given the frequency and mean flow profile, we can use PSE to compute the corresponding streamwise evolution of the instability intensity. The parameter $\epsilon_i S_0$ is obtained by fitting the intensity envelope to a Gaussian profile, while the spatial distance $X_m/S_0$ is identified at the peak amplitude location of the instability waves. 
{Furthermore, the streamwise wavenumber, $\alpha$, is extracted from the local stability theory (LST) evaluated at the nozzle lip. In effect, this implies that the effects of the jet spreading are mainly reflected in the amplitude modulation, rather than in the change of convection velocity in the present model.}
The Mach number $M_+$ is calculated from $M_-$ using
Crocco-Busemann's rule, i.e.
\begin{equation}
    M_{+}=\dfrac{M_{-}}{\sqrt{1+\dfrac{\gamma-1}{2}M_{-}^2}}\nu^{1/2}.
    \label{M_c}
\end{equation} 
The far-field sound is presented in terms of sound pressure level (SPL) defined
by 
\begin{equation}
\mathrm{SPL}=20\log_{10}\frac{|p_a|}{|p_{r}|},
\end{equation}
where $p_{r}=2\times 10^{-5}$. 

\section{Results}
\label{sec:3}
The predictions from the parametric model are presented in this section. Typical
directivity patterns and frequency spectra of the BBSAN in the far field are first
shown in Sec. \ref{sec:3.1}. Effects of the four non-dimensional parameters are
examined in Sec. \ref{sec:3.2}, following which the predictions are compared with
experimental data in Sec. \ref{sec:3.3}.
\subsection{Typical directivity patterns and frequency spectra}
\label{sec:3.1}

Typical directivity patterns of the BBSAN are shown in Fig. \ref{fig:example2}.
In an attempt to gain more physical insights into the typical BBSAN characteristics from the model, we first start with a simplified operation condition where $\mathcal{A}_i =
\mathcal{A}_{s0} = 1$ and $\epsilon_s=\sigma=0$, with a
temperature ratio $\nu = 1$. The effects of varying these parameters will be discussed in detail in Sec. \ref{sec:3.2}. {The azimuthal mode is taken to be 1.}
We choose the spatial mismatch $X_m/S_0=3$, consistent with experimental findings \citep{2020_sound_sources}. The non-dimensional instability amplitude decay rate, $\epsilon_i
S_0$, is set to $1$, meaning the instability amplitude decreases to 37\%
over one shock spacing distance. Effects of $\epsilon_i S_0$ on the directivity pattern will be examined in Sec. \ref{sec:3.2}.   

From Fig. \ref{fig:example2}(a),
it is evident that at low frequencies, the
directivity pattern shows a major lobe directed upstream and a minor lobe to the downstream direction, consistent with the
typical BBSAN directivity observed in experiments~\citep{1988Tam}. As the angular frequency $\omega$ increases, the main lobe gradually shifts downstream and becomes narrower. Fig.~\ref{fig:example2}(b) shows a similar trend: at low jet Mach numbers, a dominant upstream lobe appears, which progressively shifts downstream with increasing $M_{-}$. Moreover, a secondary downstream lobe emerges and shows an increase in intensity as $M_{-}$ rises.

Trends found in Fig. {\ref{fig:example2}} can be directly explained by Eqs.
(\ref{equ:Fk_degenrated}) and (\ref{equ:result}). Assuming $\epsilon_s = \sigma = 0$ and {neglecting $H^{(1)}_n(\omega M_+ \sin\psi / 2)$ in Eq. (\ref{equ:result})}, the maximum radiation angle $\psi_m$ can be expressed as
\begin{equation}
\psi_m = \mathrm{arccos}\frac{{\alpha} - 2\pi / S_0}{\omega M_+}, \quad \frac{{\alpha} - 2\pi / S_0}{\omega M_+} > -1.
\label{equ:3.1}
\end{equation}
{A brief discussion on the influence of $H^{(1)}_n(\omega M_+ \sin\psi / 2)$ is provided in Appendix~\ref{appendix:hankel}.} When $({\alpha} - 2\pi / S_0)/{\omega M_+} < -1$, the maximum radiation angle approaches $180^\circ$, as shown by Appendix~\ref{appendix:hankel}. Otherwise, as $\omega$ and $M_-$ increase, $\psi_m$ decreases; therefore, the major radiation lobe moves progressively towards the downstream direction.

\begin{figure}
   \centering
   \includegraphics[width = 0.65\textwidth]{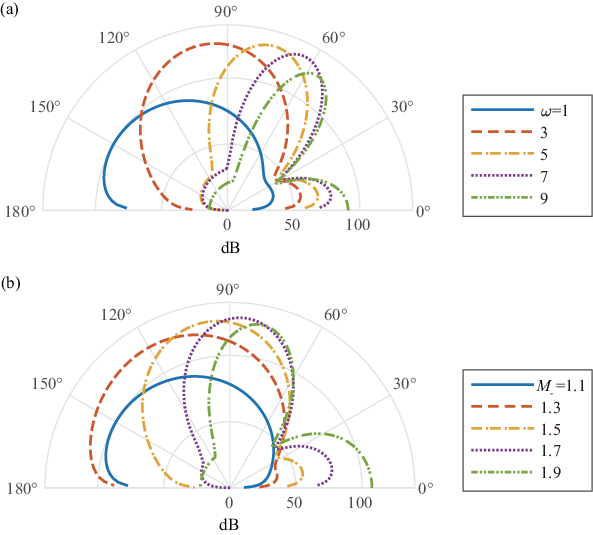}
   \caption{Typical directivity patterns of the BBSAN in the far field. The sound pressure fluctuation is calculated via Eq. (\ref{equ:result}) with $R=1$. (a) $M_{-}=1.5$; (b) $\omega=3$.}
\label{fig:example2}
\end{figure}

\begin{figure}
   \centering
   \includegraphics[width = 0.8\textwidth]{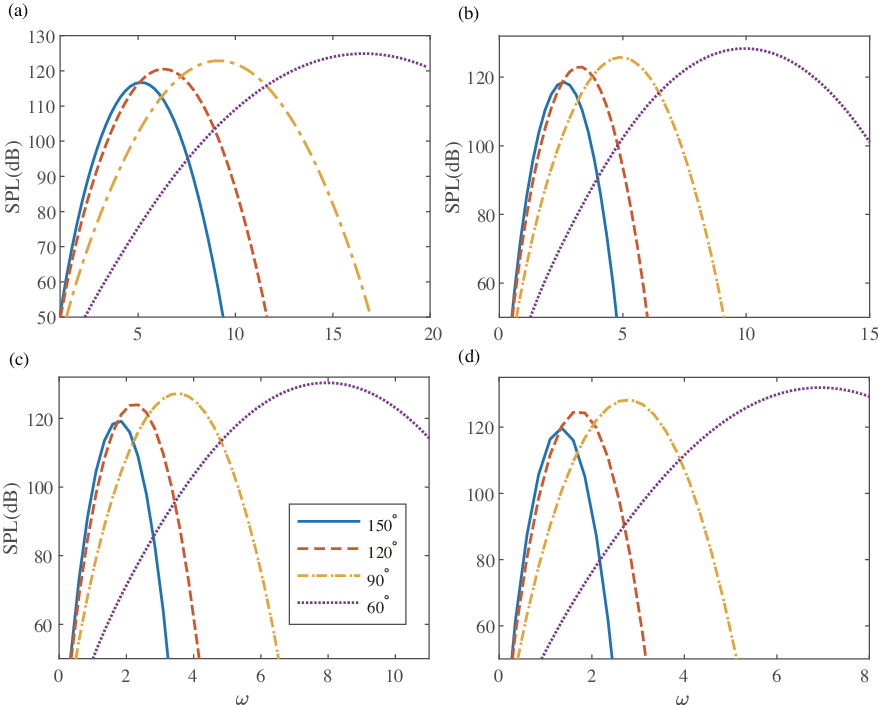}
   \caption{Typical frequency spectra of the BBSAN in the far field. The sound pressure fluctuation is calculated via Eq. (\ref{equ:result}) with $R=1$. (a) $M_{-}=1.1$; (b) $M_{-}=1.3$; (c) $M_{-}=1.5$; (d) $M_{-}=1.7$. }
\label{fig:example1}
\end{figure}

Representative frequency spectra of the acoustic waves induced by the SII are shown in Fig.~\ref{fig:example1}, under operation conditions identical to those in Fig.~\ref{fig:example2}.  
Four jet Mach numbers are considered, namely \( M_{-} = 1.1, 1.3, 1.5, 1.7 \).  
For each Mach number, spectra at four distinct observer angles are presented.  
First, the spectrum at each Mach number exhibits a distinct peak, with the peak frequency \( \omega_m \) decreasing as the observer angle \( \psi \) increases.  
Second, the bandwidth of the spectral peak increases with a decreasing observer angle.  
Third, as the jet Mach number increases, both the peak frequency \( \omega_m \) and the bandwidth decrease.

These trends can also be explained by Eq. (\ref{equ:Fk_degenrated}). The maximum frequency $\omega_m$ reads 
\begin{equation}
\omega_{m}=\frac{2\pi}{S_0(1/U_c-M_+\cos\psi) }, 
\label{equ:fCalculation}
\end{equation}
where $U_c=\alpha/\omega$ represents the convection velocity of the instability waves. The derivation of Eq. (\ref{equ:fCalculation}) is provided below. When $\psi>\arccos 1/M_{+} U_c$, the peak frequency $\omega_{m}$ increases as the observer angle $\psi$ decreases. Additionally, an increase of $M_-$ (equivalent to the increase of $M_{+}$) leads to a decrease of $\omega_{m}$.

{To estimate the bandwidth of the spectral peak, note the appearance of $X =
(\alpha + k)S_0$ in Eq.        (\ref{equ:Fk_degenrated}). Strictly speaking, both
$\epsilon_i$ and $\mathcal{A}_i$ depend on $\omega$ and an direct calculation of
the bandwidth from Eq. (\ref{equ:Fk}) appears intractable. To facilitate a
 quick understanding of the trend observed in Fig.~\ref{fig:example1}, we assume that they do not
change significantly within the frequency range of interest (as can also be
verified). Substituting $X = (\alpha + k)S_0$  into Eq. (\ref{equ:result}) with $k =
-\omega M_+\cos\psi$ yields the far‐field acoustic pressure. The acoustic
intensity reaches the maximum when $X = 2\pi$, which gives the spectral peak
frequency in Eq. (\ref{equ:fCalculation}).  
If the SPL decreases by $\Delta \mathrm{SPL}$ from its peak value, and the corresponding change in $X$ is represented by
$2\pi\delta$, then we have
\begin{equation}
\Delta \mathrm{SPL}=20\mathrm{log}_{10}\frac{p_{\mathrm{max}}}{p_\delta}\approx20\mathrm{log}_{10}\mathrm{exp}\left(\frac{\pi^2\delta^2}{\epsilon_i^2 S_0^2}\right),
\label{equ:DSPL}
\end{equation}
where $p_{\mathrm{max}}$ and $p_\delta$ represent the peak pressure amplitude and the amplitude after the $2\pi\delta$ change in $X$, respectively.
The corresponding bandwidth follows from
\begin{equation*}
2\pi\delta = ({1}/{U_c} - M_+\cos\psi)S_0\Delta\omega,
\end{equation*}
so that
\begin{equation}
\Delta\omega 
= \frac{2\pi/S_0}{({1}/{U_c} - M_+\cos\psi)} \delta.
\label{equ:this}
\end{equation}
Therefore, the
bandwidth is proportional to
\begin{equation}
\Delta \omega \sim \frac{1}{1/U_c-M_+\cos\psi }.
\label{equ:3.4}
\end{equation}}

From Eq. {(\ref{equ:3.4})}, the dependence of the spectral bandwidth on the Mach number and observer angle is clear. One can see that when $M_+$ or $\psi$ decrease {(given that $\psi>\pi/2$ and therefore $\cos\psi<0$)},
the bandwidth increases
, which explains the trend observed in Fig.
\ref{fig:example1}. However, if $\psi$  further decreases and approaches
$0^\circ$, ${1/U_c-M_+\cos\psi }$ may be less than 0, and the spectral peak disappears, {as can be observed in experiments \citep{1982NormSeiner}.}

\subsection{Effects of the non-dimensional parameters}
\label{sec:3.2}

{In this section, we perform a parametric study to examine the effects of the non-dimensional parameters on the directivity patterns and spectra, and a discussion of the underlying physical mechanisms of these effects is attempted wherever possible.} The relevant
results are shown in Figs. \ref{fig:directivity_1}-\ref{fig:spectra2}.

\begin{figure}
   \centering
   \includegraphics[width = 0.9\textwidth]{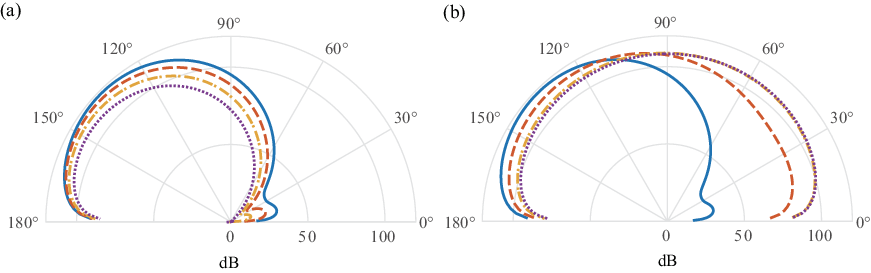}
   \caption{Effects of $X_m/S_0$ and $\epsilon_i S_0$ on the directivity pattern. The operation conditions are: $\epsilon_s S_0=10\%$ and $\sigma S_0=5\%$. (a) The jet Mach number is $M_{-}=1.5$ and the angular frequency is $\omega=2$. The parameter $X_m/S_0=3, 4, 5, 6$ for the {solid}  
   line, dashed line, dash-dotted line, and dotted line, respectively;  (b) The jet Mach number is $M_{-}=1.5$ and the angular frequency is $\omega=3$. The parameter  $\epsilon_i S_0=1, 2, 4, 6$ for the solid line, dashed line, dash-dotted line, and dotted line, respectively. }
\label{fig:directivity_1}
\end{figure}

The effects of $X_m/S_0$ and $\epsilon_i S_0$ on the directivity patterns are shown in Fig. \ref{fig:directivity_1}. The jet Mach number and angular frequency are selected to reflect typical operation conditions in experiments.
 The shock-associated parameters, i.e., $\epsilon_s S_0$ and $\sigma S_0$, are set to be 10\% and 5\%, respectively. These values fall within the experimentally observed ranges, as shown in Figs. \ref{fig:exampl3} and \ref{fig:exampl4}.  From Fig. \ref{fig:directivity_1}(a), we can see that increasing $X_m/S_0$ reduces the SPL at all observer angles, which is expected, since a larger $X_m/S_0$ means a greater spatial offset between the location of maximum instability and the shock intensity, which in turn reduces the resulting BBSAN. 

Regarding $\epsilon_i S_0$, as shown in Fig. \ref{fig:directivity_1}(b), the primary
radiation direction remains unchanged as $\epsilon_i S_0$ increases from 1 to 2, while the peak SPL value decreases and the intensity of the secondary lobe increases. However, as
$\epsilon_i S_0$ further increases, the directivity pattern gradually expands and turns into a monopole-like circular shape. {This is not surprising, as a higher  $\epsilon_i
S_0$ implies a more rapid reduction in instability intensity and a more localized effective acoustic source. From the wavenumber perspective, this leads to a broader wavenumber spectrum.}
In contrast, a wider acoustic source region {leads to a more localized
wavenumber distribution and therefore
produces a more directional acoustic emission.} 

The effects of shock amplitude decay $\epsilon_s S_0$ and shock spacing decay $\sigma S_0$ on the directivity patterns are shown in Fig. \ref{fig:directivity_2}. In Fig.~\ref{fig:directivity_2}(a), increasing $\epsilon_s S_0$ from 1/18 to 1/4 leads to a modest reduction in the peak SPL when $X_m/S_0 = 3$. This limited effect occurs because the location of maximal instability waves lies close to the nozzle exit; therefore, enlarging  $\epsilon_s S_0$ only slightly alters the SPL. In contrast, when $X_m/S_0$ is increased to 6, the cumulative impact of $\epsilon_s S_0$ becomes pronounced: the peak SPL drops significantly as $\epsilon_s S_0$ varies from 1/18 to 1/4. Similarly, increasing \(\sigma S_0\) induces an upstream shift of the peak radiation angle at \(X_m/S_0 = 3\), whereas this shift becomes more pronounced when \(X_m/S_0 = 6\) (Fig.~\ref{fig:directivity_2}(c)–(d)). This behavior can be explained by~Eq. (\ref{equ:3.1}) (given that $\sigma$ and $\epsilon_s$ are relatively small and Eq. (\ref{equ:3.1}) remains approximately valid): an increase in \(\sigma S_0\) leads to a decrease in the effective shock spacing, which causes the peak radiation angle to move upstream. 
These results demonstrate that the effects of \(\sigma S_0\) and \(\epsilon_s S_0\) on the directivity pattern depend strongly on \(X_m\). For larger values of \(X_m\), even a slight increase in \(\sigma S_0\) and \(\epsilon_s S_0\) leads to substantial changes in the directivity. Furthermore, even at lower \(X_m\), the influence of \(\sigma S_0\) remains significant and should not be overlooked. For instance, an increase in \(\sigma S_0\) from 0 to 5\% results in an SPL variation of up to 20 dB at particular observer angles. 

\begin{figure}
   \centering
   \includegraphics[width = 0.9\textwidth]{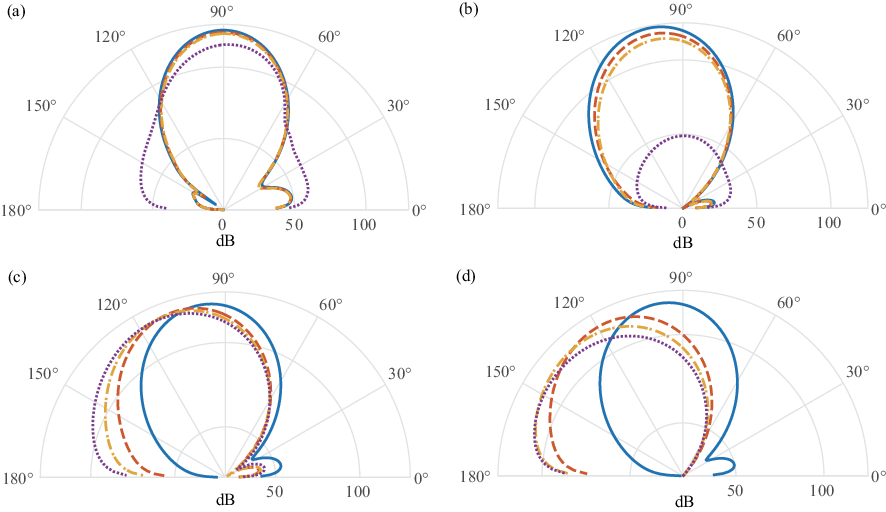}
   \caption{Effects of $\sigma S_0$ and $\epsilon_i S_0$ on the directivity pattern. The operation conditions are $\epsilon_i S_0=1$ and $M_{-}=1.5$ (a) The angular frequency $\omega=4$, $\sigma S_0=5\%$, and $X_m/S_0=3$. The parameter $\epsilon_s S_0=1/18, 1/10, 1/8, 1/4 $ for the solid line, dashed line, dash-dotted line, and dotted line, respectively;  (b) Operation conditions are the same as (a) except that $X_m/S_0=6$; (c)  The angular frequency $\omega=3$, $\epsilon_s S_0=10\%$, and $X_m/S_0=3$. The parameter $\sigma S_0=0, 1/20, 1/15, 1/12$ for the solid line, dashed line, dash-dotted line, and dotted line, respectively; (d)  Operation conditions are the same as (c) except that $X_m/S_0=6$.}
\label{fig:directivity_2}
\end{figure}

\begin{table}
    \begin{center}
    \begin{tabular}{lc}
    \toprule
        Parameter   & Effect   \\[3pt]
        $X_m/S_0$ & Reduces the SPL \\  
           $\epsilon_s S_0$ & Reduces the SPL \\  
        $\epsilon_i S_0$   & Reduces the peak SPL and broadens the directivity pattern \\  
        $\sigma S_0$ & Shifts the directivity patterns to the upstream direction \\  
    \end{tabular}
    \caption{{Effects of increasing the non-dimensional parameters on the directivity pattern.}}
    \label{tab:2}
    \end{center}
\end{table}

\begin{figure}
   \centering
   \includegraphics[width = 0.9\textwidth]{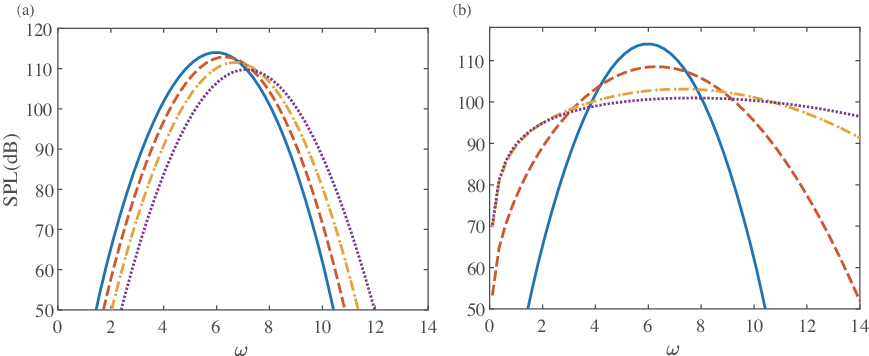}
   \caption{Effects of $X_m/ S_0$ and $\epsilon_i S_0$ on the directivity pattern. The opreation conditions are $M_{-}=1.1$ $\psi=150^\circ$. Other operation conditions for (a) and (b) are the same as those in Figs. \ref{fig:directivity_1}(a, b), respectively.}  
\label{fig:spectra1}
\end{figure}

The effects of these four non-dimensional parameters on the directivity patterns are summarized in Table \ref{tab:2}. Their influence on the frequency spectra can be examined in a similar manner. The operation conditions are set to $M_- = 1.1$, and the observer angle is fixed at $150^\circ$. Trends at other observer angles are similar and, therefore, not presented here for brevity. All other operation conditions remain {identical} with those in Figs. \ref{fig:directivity_1} and \ref{fig:directivity_2}.
As shown in Fig. \ref{fig:spectra1}(a), increasing $X_m/S_0$ results in a reduction of the peak SPL while simultaneously shifting the spectra slightly toward higher frequencies. Although Eq. (\ref{equ:fCalculation}) is only strictly valid for vanishing $\epsilon_s$ and $\sigma$, considering they are both of small amplitudes, we might still use it to explain the behavior observed here.
An increase in $X_m/S_0$ leads to a decrease in the shock spacing around the effective source position. Consequently, the peak frequency $\omega_m$ increases.
From Fig.~\ref{fig:spectra1}(b), increasing $ \epsilon_i S_0 $ reduces the peak SPL and broadens the spectra. {The reduction in peak SPL is expected,  since as $\epsilon_i S_0$ grows, the effective shock strength contributing to the SII also decreases due to the rapid attenuation of the instability waves.} The change in the spectral bandwidth can be approximately interpreted via~Eq. (\ref{equ:DSPL}) given that $\epsilon_s$ and $\sigma$ are of small amplitude. It shows that the SPL variation near the spectral peak diminishes as $ \epsilon_i S_0 $ increases, thereby increasing the bandwidth.
\begin{figure}
   \centering
   \includegraphics[width = 0.9\textwidth]{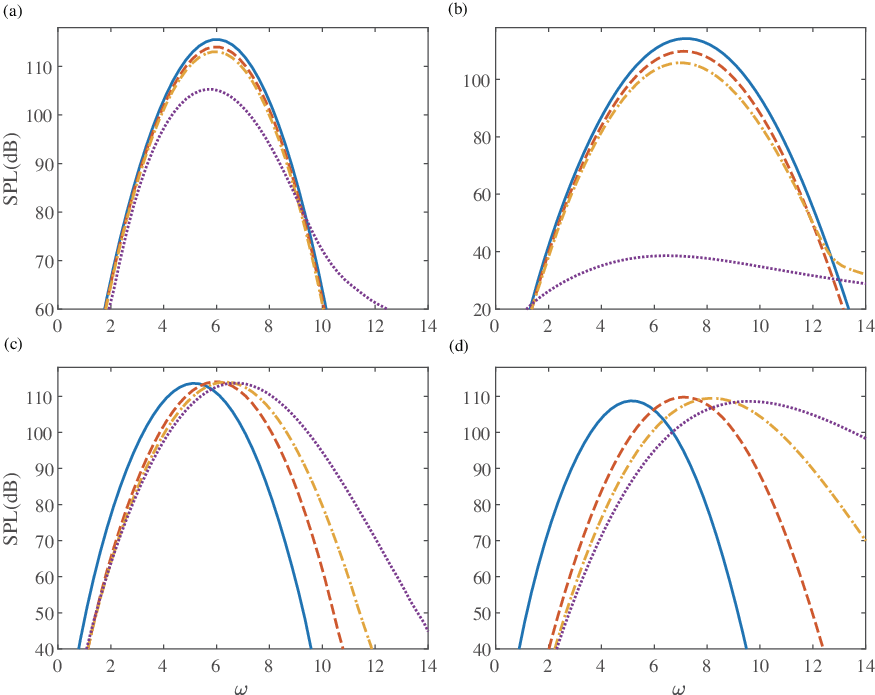}
   \caption{Effects of $\sigma S_0$ and $\epsilon_s S_0$ on the directivity pattern. The operation conditions are $M_{-}=1.1$ $\psi=150^\circ$.  Other operation conditions for (a, b, c, d) are the same as those in Figs. \ref{fig:directivity_2}(a, b, c, d), respectively.}
\label{fig:spectra2}
\end{figure}

\begin{table}
    \begin{center}
    \begin{tabular}{lc}
    \toprule
        Parameter   & Effect  \\[3pt]
        $X_m/S_0$ & Reduces the peak SPL and shifts the spectra toward higher frequencies\\ 
        $\epsilon_i S_0$   &  Reduces the peak SPL and broadens the spectrum \\  
        $\epsilon_s S$ & Reduces the peak SPL  \\
        $\sigma S_0$ & Shifts the spectra toward higher frequencies \\ 
    \end{tabular}
    \caption{{Effects of increasing the non-dimensional parameters on the frequency spectrum.}}
    \label{tab:3}
    \end{center}
\end{table}
Similar to Figs.~\ref{fig:directivity_2}(a, b), Figs.~\ref{fig:spectra2}(a, b) demonstrate that raising $ \epsilon_s S_0 $ reduces the peak SPL, with the magnitude of this reduction growing for larger values of $ X_m/S_0 $. {In addition, an increase in $ \sigma S_0 $ shifts the spectra to higher frequencies and increases the spectral bandwidth. Both effects become more pronounced as $X_m/S_0$ increases. }   
 The effects of these four non-dimensional parameters on the frequency spectra are summarized in Table \ref{tab:3}.

\subsection{Comparison with experimental data}
\label{sec:3.3}

To validate the present model, this section compares its predictions with experimental measurements reported by \citet{1982NormSeiner} and \citet{Yu_1972}. The comparisons for the frequency spectra and the directivity patterns are presented separately.

{Regarding the frequency spectra, we first compare the predicted full spectra with the experimental data \citep{1982NormSeiner}. The operation conditions are $M_d=1$ and $\beta=1$. {Since at relatively high Mach numbers (e.g., $M_{-} > 1.3$) the instability waves in the helical and flapping modes become more dominant than those in the axisymmetric mode \citep{2019EDINGTON}, the azimuthal mode is therefore set to $n = 1$ for $M_{-} > 1.3$ in the subsequent analysis.} Shock-associated parameters are obtained from the measured static pressure {by \citet{1982NormSeiner}} on the jet centre line; the resulting values are $\sigma S_0=5\%$ and $\epsilon_s S_0=8.6\%$.} 

As introduced in Sec. \ref{subsection:2.2}, the frequency-dependent $\epsilon_i S_0$ and $\mathcal{A}_i$ are determined by linear stability analysis based on the PSE. From the measured static pressure in the jet center line, we can find that the potential core length is around 9, which agrees well with that predicted by the empirical formula (\ref{equ:empirical_formula}). The PSE is then initiated using the mean flow profile reconstructed by the potential core length. {An outline of this reconstruction procedure is shown in Appendix \ref{appendix:a}.} Further details of the PSE method can be found in \citet{PSE_piot}.} {To put this into perspective, Fig. \ref{fig:example444}(a) shows the obtained $A_i$ and $\epsilon_i S_0$ with respect to frequency using the PSE. Figure~\ref{fig:example444}(a) shows that the instability wave amplitude $\mathcal{A}_i$ (normalized by its maximum value) initially increases with $\omega$, reaches a maximum near $\omega = 1.5$, and then decreases as $\omega$ continues to increase. Figure \ref{fig:example444}(b) shows that $\epsilon_i S_0$ increases monotonically with $\omega$. Both trends are consistent with experimental findings.

\begin{figure}
   \centering
   \includegraphics[width = 0.9\textwidth]{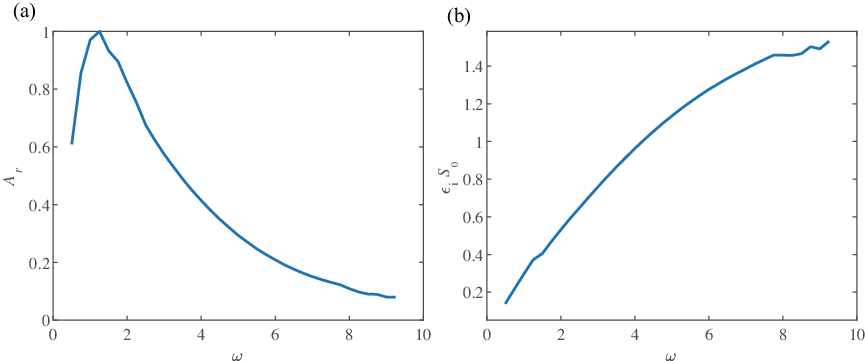}
   \caption{{Results calculated from the PSE under the operation condition $M_d=1$ and $\beta=1$. (a) The normalized $\mathcal{A}_i$ with respect to its maximum value. (a) The decay rate of instability waves $\epsilon_i S_0$. }}
\label{fig:example444}
\end{figure}

\begin{figure}
   \centering
   \includegraphics[width = 0.95\textwidth]{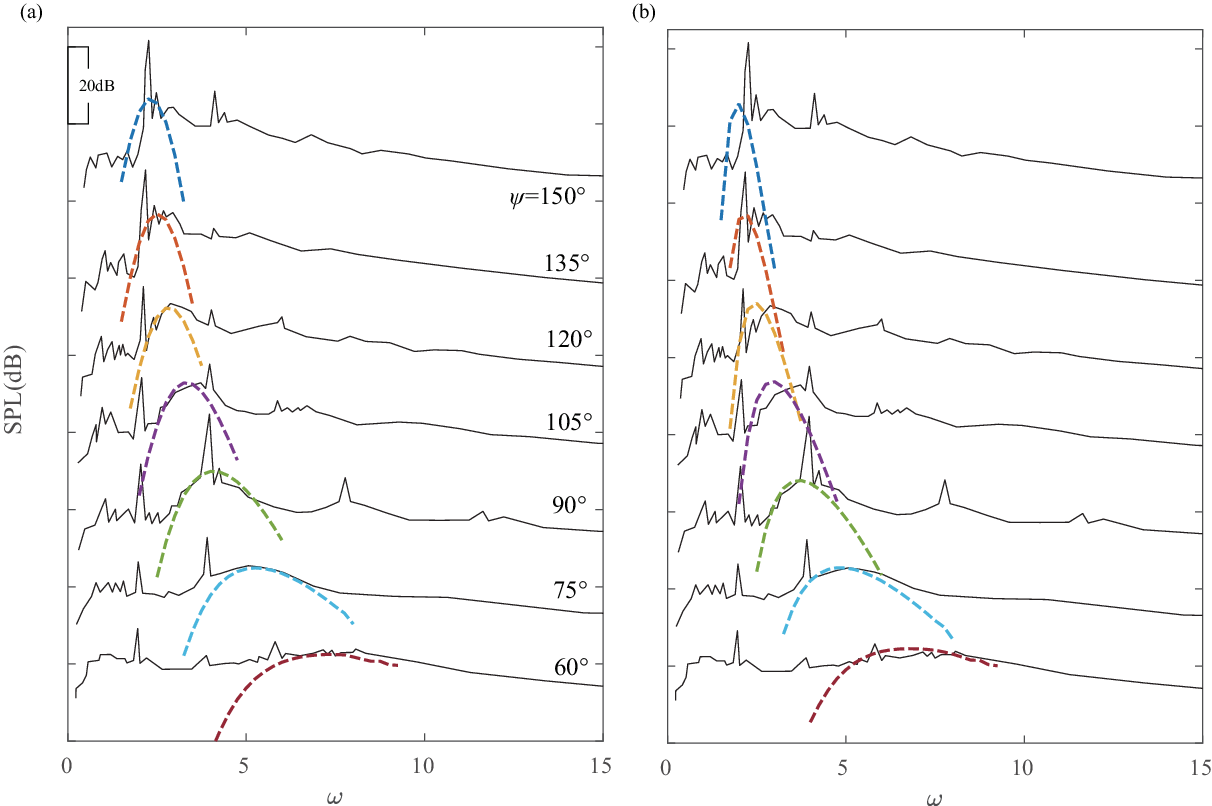}
   \caption{{Comparison of the predicted full spectra with the experimental data \citep{1982NormSeiner}. The dashed and solid lines represent the predictions and experimental data, respectively. (a) The parameters $\epsilon_s S_0$ and $\sigma S_0$ are determined via experimental data ($\epsilon_s S_0=8.6\%, \sigma S_0=5\%$); (b) $\epsilon_s S_0=0\%$ and $\sigma S_0=0\%$. The instability-related parameters $\epsilon_i$ and $\mathcal{A}_i$ are calculated using PSE.  The amplitude of the predicted SPL is adjusted to match the experimental data.}}
\label{fig:realEA}
\end{figure}
{Figure \ref{fig:realEA} shows the
comparison between predicted and measured sound spectra at various observer
angles. Note that the amplitudes of predictions at each observer angle are adjusted to match with the experimental data. One can see from Fig.~\ref{fig:realEA}(a) that the predictions from this model accurately capture the spectral peak induced by the BBSAN. At small
angles
(e.g., $\psi=60^\circ$), this model captures a 
broad peak centered around $\omega\approx 7$, in agreement with the measurements.
As $\psi$ increases toward $90^\circ$ and beyond, the predicted peak shifts to
lower frequencies and its bandwidth reduces—closely matching the measured
spectral contraction. 
These results demonstrate that the present
semi-analytical model captures both the peak location and bandwidth variation of
the BBSAN spectra across a wide range of
observer angles.}

The bandwidth at various observer angles may be quantitatively compared. 
From the experiment, the spectral peak height exceeds 10 dB when $\psi = 150^\circ$, but drops to below 5 dB when $\psi = 60^\circ$. A similar trend is also observed in Fig.~\ref{fig:spectra1}(b) as $\epsilon_i S_0$ increases. This reduction in spectral peak height can also be explained by Eq. (\ref{equ:DSPL}). When
$\psi$ decreases, the spectral peak shifts to higher frequencies, causing the
corresponding $\epsilon_i S_0$ to increase, as shown in Fig.
\ref{fig:example444}(b). Consequently, if $\delta$ remains virtually unchanged
(as shown by Fig. \ref{fig:example4}), the bandwidth becomes larger.
{A quantitative comparison on the spectral full bandwidth is provided in Appendix
\ref{appendix:bandwidth}.

{To quantify the effects of including the spacing and amplitude decay of the
shock structures, Fig.~\ref{fig:realEA}(b) shows the predicted spectral peaks
when uniform shock structures are assumed, i.e. both $\epsilon_s S_0$  and $\sigma
S_0$ are set to be 0. As shown in Fig. \ref{fig:realEA}(b), the predicted
spectral peaks agree satisfactorily at low observable angles; however, the
discrepancies become increasingly pronounced when the observer angle
increases. For example, when the observer angle is beyond $105^\circ$, a clear
under-prediction of the peak frequency occurs, and the predicted bandwidth also
appears narrower than that measured in experiments. These discrepancies become
even more pronounced when $X_m/S_0$ is set to zero (results not shown for
brevity). Figure \ref{fig:realEA}(b) further highlights the importance of
accounting for the variation due to $\epsilon_s S_0$ and $\sigma S_0$ in the
spectral prediction of the BBSAN.}

In addition to the spectra, the predicted directivity patterns of the BBSAN are
also compared with the experimental data. Note that experiments measured the
overall jet noise, which included various noise components such as the turbulent
mixing noise (TMN) and BBSAN. However, this model only predicts the BBSAN.
{To exclude the TMN from the total jet noise measured in experiments,
considering that TMN is generated due to turbulence and is not associated with
shock structures, we use the following method. First, at the design condition of
the jet, the acoustic pressure due to TMN
$p_{\mathrm{TMN_0}}$ is estimated via
\begin{equation*}
p_{\mathrm{TMN_0}}=10^{\frac{\mathrm{SPL}_{\mathrm{Base}}}{20}}p_r,
\label{equ_TMNSPL}
\end{equation*}
where $\mathrm{SPL}_{\mathrm{Base}}$
denotes the SPL measured under design conditions. Under the design condition, no
shock structures are present, therefore, we assume that only the turbulent mixing noise
contributes to the far-field sound measurement. 

{The TMN under the off-design condition $p_{\mathrm{TMN_1}}$ can be then
estimated from $p_{\mathrm{TMN_0}}$ via Lighthill's acoustic analogy
\citep{1952lighthill}, i.e.
\begin{equation}
p_{\mathrm{TMN_1}}=\frac{M_{-}^4}{M_d^4}\frac{\tilde{D}_j}{\tilde{D}} p_{\mathrm{TMN_0}},
\label{equ:estimate}
\end{equation}
where the ratio ${\tilde{D}_j}/{\tilde{D}}$ may be calculated following the
method proposed by \citet{1988Tam_shockspacing}.} Equation (\ref{equ:estimate})
may not be very accurate, but it provides a reasonable estimate of the TMN that is not possible to measure separately in experiments. With this
estimation, the BBSAN-induced pressure perturbation can be calculated using
$p_{\mathrm{BBSAN}}=p_{\mathrm{Total}}-p_{\mathrm{TMN_1}}$, and one obtains
\begin{equation}
\mathrm{SPL}_{\mathrm{BBSAN}}=20\mathrm{log}_{10}\frac{|p_{\mathrm{BBSAN}}|}{p_r}.
\label{equ_BBSANSPL}
\end{equation}
Here the total acoustic pressure $p_{\mathrm{Total}}$ is calculated from
$\mathrm{SPL}_{\mathrm{Total}}$, i.e.
\begin{equation*}
p_{\mathrm{Total}}=10^{\frac{\mathrm{SPL}_{\mathrm{Total}}}{20}}p_r.
\label{equ_total}
\end{equation*}
Note that although Eq. (\ref{equ_BBSANSPL}) provides a
reasonable estimation of the BBASN from the total noise in the general case, the
error is likely to be significant when $p_{\mathrm{TMN_1}}$ is close to
$p_{\mathrm{Total}}$. This is likely the case at low observer angles ($\psi\approx 30^\circ$), where the TMN
is particularly strong due to large coherent structures.

We first compare the model predictions with the experimental data reported by
\citet{Yu_1972}, where $M_{-} = 1.6$ and the designed Mach number is $M_d =
1.5$. In the experiments, directivity patterns for both $M_{-} = 1.6$ and $M_{-}
= 1.5$ were measured, allowing $\mathrm{SPL}_{\mathrm{BBSAN}}$ to be estimated
using Eq. (\ref{equ_BBSANSPL}). The parameters $\epsilon_s S_0$, $\sigma S_0$, and
the potential core length can be determined from the measured centerline Mach
number distribution (Fig. 3 in \citet{Yu_1972}). {The potential core length
is around 10, agreeing well with that predicted by
Eq. (\ref{equ:empirical_formula}).} Parameters $X_m/S_0$,
$\epsilon_i S_0$, and $\mathcal{A}_i$ can then be calculated using the PSE method. Note that the
intensity of the predicted directivity patterns is scaled to match the
experimental data, however, the scaling is calibrated at one single frequency, but remains the same for all other frequencies shown in this section. 

\begin{figure}
   \centering
   \includegraphics[width = 0.95\textwidth]{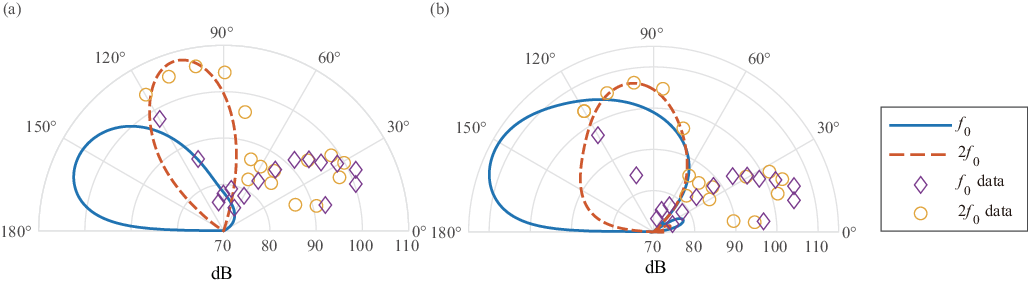}
   \caption{Comparisons of the predicted directivity patterns with experimental data \citep{Yu_1972}. The jet Mach number is $M_{-}=1.6$ and the frequency is $f_0=0.23$. (a) Parameters for the instability wave at $f=f_0$ and $f=2f_0$ are respectively $X_m/S_0=5.32, 2.43$, $\epsilon_i S_0=0.37, 0.76$, and $A_r=1, 0.57$. {The shock-associated parameters are $\epsilon_s S_0=9.6\%$ and $\sigma S_0=4.5\%$; (b) Three parameters, i.e., $X_m/S_0$, $\epsilon_s S_0$, and $\sigma S_0$ are set to zero.} }
\label{fig:ju}
\end{figure}

As shown in Fig. \ref{fig:ju}(a), the far-field directivity patterns of the BBSAN
at two frequencies are presented.  We can see that at $f_0$ the predicted
maximal radiation angle $\psi$ is larger than $120^\circ$, while at $2f_0$ it is
around $100^\circ$. The predictions agree well with the experimental data in
terms of the maximal radiation direction at $2f_0$; although the maximal angle is
not known in experiments due to limited data at $f_0$, the predicted shape does
follow the experiment rather closely. However, in the downstream direction,
a pronounced lobe is observed for both frequencies, which is not present in the
predictions. This is likely due to a failed estimation of the BBSAN using Eq.
(\ref{equ_BBSANSPL}) because of the strong mixing noise at low observer angles.

{When the three shock-associated parameters, $X_m/S_0$, $\epsilon_s S_0$, and $\sigma S_0$, are set to zero, as shown in Fig. \ref{fig:ju}(b), the prediction at $2f_0$ remains in good agreement with the experimental data. In contrast, the prediction at $f_0$ deviates, with the SPL pronouncedly overpredicted for $\psi>60^\circ$. This demonstrates the importance of accounting for variations associated with the shock structures in predicting the directivity of BBSAN.}

\begin{figure}
   \centering
   \includegraphics[width =0.95\textwidth]{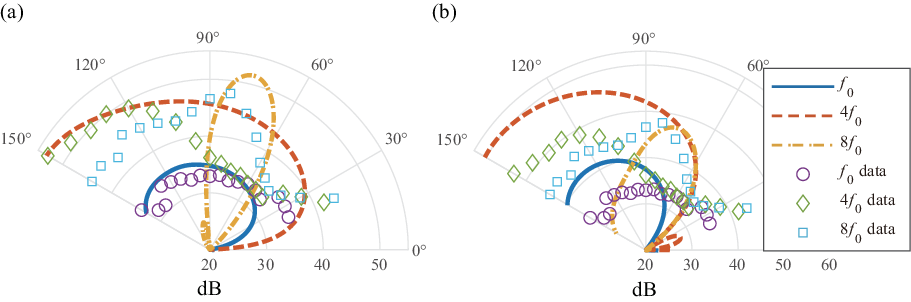}
   \caption{Comparisons of the predicted directivity patterns with experimental
   data \citep{1982NormSeiner}. The parameters $\epsilon_s S_0$ and $\sigma S_0$ are determined from the experimental data (shown in Figs. \ref{fig:exampl3} and \ref{fig:exampl4}), while $\epsilon_i S_0$, $A_r$, and $X_m/S_0$ are calculated from the PSE. The operation conditions are (a) $M_{-}=1.49$, $[X_m/S_0, \epsilon_i S_0, A_r]$ $=[6.43, 0.21, 2.97], [3.72, 0.37, 3.86]$, and [1.00, 1.11, 1] for solid, dashed, and dash-dotted lines, respectively.  The shock-associated parameters are $\epsilon_s S_0=10\%$ and $\sigma S_0=3.8\%$; (b) Operation conditions are the same as (a) except that three parameters, i.e. $X_m/S_0$, $\epsilon_i S_0$, and $\sigma S_0$ are set to zero.} 
\label{fig:example5}
\end{figure}
{To further validate the model, we compare the predicted directivity} with the experimental data measured by
\citet{1982NormSeiner}, as shown in Fig. \ref{fig:example5}. The designed jet
Mach number is $M_d = 1$, and directivity patterns at $M_{-} = 1$ were also
measured in \citet{1982NormSeiner}. Thus, the TMN is excluded
using Eq. (\ref{equ_BBSANSPL}). The
dimensional $\tilde{f}_0$ in Fig. \ref{fig:example5} is 1000 Hz, while
the non-dimensionalized frequency is 0.104.

{As shown in Fig.~\ref{fig:example5}(a), the measured directivity pattern at $f_0$ exhibits an approximately monopole-like distribution. This is likely because $f_0$ lies within the frequency range where the TMN component is considerably stronger than the BBSAN \citep{1995_annu_tam}. 
Consequently, the characteristic directivity of BBSAN is buried due to the small signal-to-noise ratios. 
In contrast, the predicted directivity exhibits a major lobe in the upstream direction that matches the experimental data well. At $4f_0$, a pronounced upstream lobe is observed in experiments. Although not exactly collapsing, the predicted lobe shape, in particular its amplitude, agrees well with the experimental data. As the frequency increases to $8f_0$, the experimental data reveal an additional lobe emerging near $\psi = 90^\circ$, where the SPL increases by approximately 10 dB relative to that at $f_0$. The model successfully captures both the emergence of this new lobe and the associated SPL enhancement.}
{However, at $8f_0$, {this model only predicts the major lobe}, while the sound radiation in other directions, particularly when
the observer angle exceeds the maximum radiation angle, is underpredicted.
This could be due to the fact that, in real jets, in
addition to large-scale instability waves, fine-scale turbulence might interact with shock structures to produce additional sound. These additional acoustic waves are not considered in the current model.}

{When the three parameters, $X_m/S_0$, $\epsilon_s S_0$, and $\sigma S_0$, are set to zero, as shown in Fig. \ref{fig:example5}(b), the predictions at $8f_0$ remain in good agreement with the experimental data in terms of the major lobe. In contrast, the predictions at $f_0$ and $4f_0$ deviate noticeably in the magnitude and general shape. This further highlights the importance of accounting for variations associated with shock structures in predicting the directivity of BBSAN.}

\section{Conclusion}
\label{sec:4}
A semi‑empirical model is developed in this paper to predict the broadband
shock‑associated noise (BBSAN) in supersonic jets, with particular emphasis on
modeling the shock and instability structures as realistically as possible. The
model integrates a modified form of Pack’s model, which accounts for the
downstream decay of both  the shock amplitude and spacing, with a wave‑packet
representing instability waves involved in the shock–instability interaction
(SII). The SII is modeled semi-empirically as a simple product of the pressure fluctuations
induced by the shock and instability waves, which is used as a boundary
condition of the Helmholtz equation on the jet lip line to calculate the
far‑field acoustic pressure.

This model successfully captures several characteristic BBSAN features observed
in the experiments, including upstream-directed radiation lobes in the
directivity patterns at low frequencies, which shift downstream with increasing
frequency or Mach number. Spectral predictions show distinct peaks whose
frequency decreases when increasing observer angles, accompanied by spectral
broadening at smaller observer angles. Parametric analysis reveals that
increasing the spatial offset between shock and instability maxima ($X_m/S_0$)
reduces the SPL while shifting spectra toward higher frequencies. Increasing the
non-dimensional instability decay rate ($\epsilon_i S_0$) significantly changes
the spectral shape, reduces the peak SPL, and broadens both the spectra and directivity
patterns. Increasing the non-dimensional shock amplitude decay rate ($\epsilon_s S_0$) leads to a decreasing SPL, which is more pronounced at larger
spatial offsets $X_m$, while increasing the non-dimensional shock spacing decay
rate ($\sigma S_0$) shifts the main acoustic radiation angle towards upstream
and spectral {peak} to higher frequencies.

Validation against multiple experimental datasets demonstrates that the model
can predict several spectral features correctly. Directivity comparisons also
show a good prediction of the lobe position; in addition, it also appears to
capture the downstream shifts of the main radiation lobe as the frequency
increases. {When the shock-associated parameters, such as intensity and spacing decay rate, are neglected, the predictions show poorer agreement with the experimental data for both spectra and directivity patterns. This highlights that incorporating more realistic representations of the shock and instability waves is important in an accurate prediction of BBSAN's spectra and directivity, which is hoped to help gain further insight into the noise physics.} 

{Note that the present work neglects the effects of engine nozzles on the generation and propagation of BBSAN; future work includes studying BBSAN under the scattering of engine nozzles or nearby wings, which is known to change the characteristics of jet noise significantly in subsonic regimes \citep{lyu2017,lyu2018,Lyu_Dowling_2019}.}

\section*{Acknowledgments}
The authors wish to gratefully acknowledge the National Natural Science Foundation of China (NSFC) under the grant number 12472263. The second author (BL) wishes to acknowledge the funding from the Beijing Natural Science Foundation (L253027) and from Laoshan Laboratory (LSKJ202202000).


\appendix
\section{}
\label{appendix:hankel}

In this section, we examine the validity of neglecting $H^{(1)}_n(\omega M_+ \sin\psi / 2)$ in the derivation of Eqs. (\ref{equ:3.1}) and (\ref{equ:fCalculation}). As shown in Fig.~\ref{fig:hankel}(a), the directivity patterns calculated from Eq. (\ref{equ:result}), with and without considering $H^{(1)}_n(\omega M_+ \sin\psi / 2)$, exhibit similar overall shapes. When the term $H^{(1)}_n(\omega M_+ \sin\psi / 2)$ is included, however, the SPL increases slightly near $\psi \approx 90^\circ$ and decreases near $\psi \approx 0^\circ$ or $180^\circ$. The spectra exhibit similar behavior. As $\omega$ increases, the SPL increases, whereas the maximum frequency $\omega_m$ remains nearly unchanged. 

Similar trends are observed under other operation conditions and are therefore omitted for brevity. These results confirm that the maximum radiation angle and maximum frequency can be reliably evaluated using Eqs. (\ref{equ:3.1}) and (\ref{equ:fCalculation}), respectively.
\begin{figure}
   \centering
   \includegraphics[width = 0.8\textwidth]{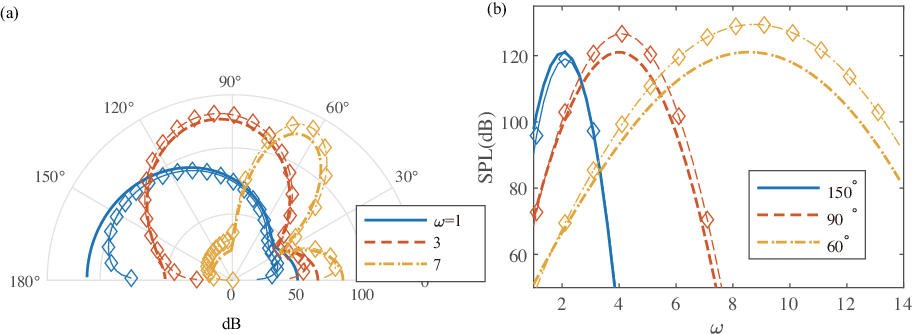}
   \caption{Effects of neglecting $H^{(1)}_n(\omega M_+ \sin\psi / 2)$ in Eq. (\ref{equ:result}) on the directivity patterns and spectra. The jet Mach number is $M_-=1.5$. Lines with markers are calculated from Eq. (\ref{equ:result}), while lines without markers are obtained by neglecting $H^{(1)}_n(\omega M_+ \sin\psi / 2)$.}
\label{fig:hankel}
\end{figure}

\section{}
\label{appendix:a}
We adopt a self-similar velocity profile, as proposed by \citet{Tam_Burton_1984}, to reconstruct the mean flow based on the length of the potential core. The jet is divided along the streamwise direction into three distinct regions: the core, a transitional region, and the fully developed mixing layer.

In the core region, the streamwise mean velocity \( \bar{u}_x \) is given by
\begin{equation}
    \bar{u}_x = \begin{cases} 
    1, & r \leq h(x), \\ 
    \exp \left\{-\ln 2\left[(r - h(x)) / b(x)\right]^2\right\}, & r > h(x), 
    \end{cases}
    \label{equ:potential}
\end{equation}
where \( h(x) \) represents the radius of the potential core and \( b(x) \) is the half-width of the mixing layer. To ensure continuity of both the velocity and its axial derivative, a transitional zone is introduced at the end of the potential core. In this region, the streamwise velocity profile is expressed as
\begin{equation}
    \bar{u}_x = W_c(x) \exp \left\{-\ln 2\left[(r - h(x)) / b(x)\right]^2\right\},
    \label{equ:transition}
\end{equation}
where \( W_c(x) \) is the streamwise velocity at the jet centerline.  
After this transition region, the core radius \( h(x) \) approaches zero, and the mean velocity simplifies to
\begin{equation}
    \bar{u}_x = W_c(x) \exp \left\{-\ln 2\left[r / b(x)\right]^2\right\}.
    \label{equ:mixing}
\end{equation}
Using the potential core length \( x_h \), the half-width of the mixing layer \( b(x) \), and the centerline velocity \( W_c(x) \), these quantities can be determined from momentum conservation in the streamwise direction, as well as from the continuity of \( h(x) \), \( b(x) \), \( W_c(x) \), and their respective derivatives with respect to \( x \).  
Once \( \bar{u}_x \) is known, the mean density \( \bar{\rho} \) can be obtained from the Crocco-Busemann's relation, and the radial velocity \( \bar{u}_r(r, x) \) is then calculated using mass conservation.

\begin{figure}
   \centering
   \includegraphics[width = 0.85\textwidth]{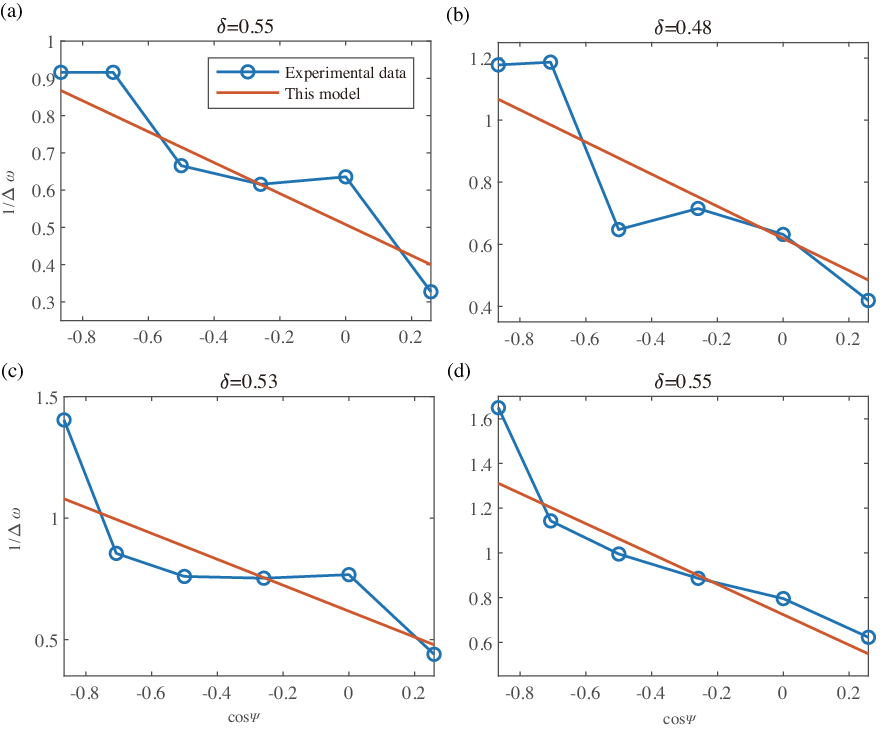}
   \caption{Comparison of predictions with experimental data \citep{1982NormSeiner} on the {full bandwidth} of frequency spectra. (a) $M_{-}=1.37$; (b) $M_{-}=1.41$; (c) $M_{-}=1.48$; (d) $M_{-}=1.67$.}
\label{fig:example4} 
\end{figure}

\section{}
\label{appendix:bandwidth}
In this section, we compare the full bandwidth of the spectral
peak, which is given by Eq. (\ref{equ:this}), with the experimental data \citep{1982NormSeiner}. The designed Mach number of the nozzle is $M_d = 1$. It
is important to note that when $M_{-}$ is relatively low, although shock
structures are clearly visible in the experiments, the spectral peaks in the
BBSAN spectra are not pronounced enough for an accurate determination of the
spectral bandwidth. This is likely due to the relatively low shock intensity, reflected by a small value of 
$|M_{-}^2 - M_d^2|$.  Therefore, we only consider four jet Mach numbers where
the spectral peaks are clearly visible and the bandwidth
can be easily determined.

As shown in Fig. \ref{fig:example4}, the full bandwidth was defined as the frequency range over which the BBSAN SPL drops by 3 dB from its peak. The solid line
represents the prediction from this model (based on Eq. (\ref{equ:this})), while the coefficient, $\delta$
in Eq. (\ref{equ:this}) is determined to provide
the best fit with the experimental data. We can see that this model reproduces the measured bandwidth well.  Interestingly, the parameter $\delta$ in Eq. (\ref{equ:this})  remains nearly
the fixed as $M_{-}$ increases. Comparison with additional experimental data measured at
$M_d = 1.5$ shows similar trends; those results are omitted here for brevity. 

\bibliography{apssamp}

\begin{thebibliography}{53}%
\makeatletter
\providecommand \@ifxundefined [1]{%
 \@ifx{#1\undefined}
}%
\providecommand \@ifnum [1]{%
 \ifnum #1\expandafter \@firstoftwo
 \else \expandafter \@secondoftwo
 \fi
}%
\providecommand \@ifx [1]{%
 \ifx #1\expandafter \@firstoftwo
 \else \expandafter \@secondoftwo
 \fi
}%
\providecommand \natexlab [1]{#1}%
\providecommand \enquote  [1]{``#1''}%
\providecommand \bibnamefont  [1]{#1}%
\providecommand \bibfnamefont [1]{#1}%
\providecommand \citenamefont [1]{#1}%
\providecommand \href@noop [0]{\@secondoftwo}%
\providecommand \href [0]{\begingroup \@sanitize@url \@href}%
\providecommand \@href[1]{\@@startlink{#1}\@@href}%
\providecommand \@@href[1]{\endgroup#1\@@endlink}%
\providecommand \@sanitize@url [0]{\catcode `\\12\catcode `\$12\catcode
  `\&12\catcode `\#12\catcode `\^12\catcode `\_12\catcode `\%12\relax}%
\providecommand \@@startlink[1]{}%
\providecommand \@@endlink[0]{}%
\providecommand \url  [0]{\begingroup\@sanitize@url \@url }%
\providecommand \@url [1]{\endgroup\@href {#1}{\urlprefix }}%
\providecommand \urlprefix  [0]{URL }%
\providecommand \Eprint [0]{\href }%
\providecommand \doibase [0]{https://doi.org/}%
\providecommand \selectlanguage [0]{\@gobble}%
\providecommand \bibinfo  [0]{\@secondoftwo}%
\providecommand \bibfield  [0]{\@secondoftwo}%
\providecommand \translation [1]{[#1]}%
\providecommand \BibitemOpen [0]{}%
\providecommand \bibitemStop [0]{}%
\providecommand \bibitemNoStop [0]{.\EOS\space}%
\providecommand \EOS [0]{\spacefactor3000\relax}%
\providecommand \BibitemShut  [1]{\csname bibitem#1\endcsname}%
\let\auto@bib@innerbib\@empty
\bibitem [{\citenamefont {Takahashi}\ \emph {et~al.}(2023)\citenamefont
  {Takahashi}, \citenamefont {Griffin},\ and\ \citenamefont
  {Grandhi}}]{highAircraft}%
  \BibitemOpen
  \bibfield  {author} {\bibinfo {author} {\bibfnamefont {T.~T.}\ \bibnamefont
  {Takahashi}}, \bibinfo {author} {\bibfnamefont {J.~A.}\ \bibnamefont
  {Griffin}},\ and\ \bibinfo {author} {\bibfnamefont {R.~V.}\ \bibnamefont
  {Grandhi}},\ }\bibfield  {title} {\bibinfo {title} {{A Review of High-Speed
  Aircraft Stability and Control Challenges.}},\ }in\ \href@noop {} {\emph
  {\bibinfo {booktitle} {AIAA AVIATION 2023 Forum AIAA Paper 23-3231}}}\
  (\bibinfo {year} {2023})\BibitemShut {NoStop}%
\bibitem [{\citenamefont {Andre}\ \emph {et~al.}(2013)\citenamefont {Andre},
  \citenamefont {Castelain},\ and\ \citenamefont {Bailly}}]{broad_band_2013}%
  \BibitemOpen
  \bibfield  {author} {\bibinfo {author} {\bibfnamefont {B.}~\bibnamefont
  {Andre}}, \bibinfo {author} {\bibfnamefont {T.}~\bibnamefont {Castelain}},\
  and\ \bibinfo {author} {\bibfnamefont {C.}~\bibnamefont {Bailly}},\
  }\bibfield  {title} {{\selectlanguage {English}\bibinfo {title} {Broadband
  shock-associated noise in screeching and non-screeching underexpanded
  supersonic jets.}},\ }\href@noop {} {\bibfield  {journal} {\bibinfo
  {journal} {AIAA J.}\ }\textbf {\bibinfo {volume} {51(3)}},\ \bibinfo {pages}
  {665} (\bibinfo {year} {2013})}\BibitemShut {NoStop}%
\bibitem [{\citenamefont {Tam}(1995)}]{1995_annu_tam}%
  \BibitemOpen
  \bibfield  {author} {\bibinfo {author} {\bibfnamefont {C.~K.~W.}\
  \bibnamefont {Tam}},\ }\bibfield  {title} {{\selectlanguage {English}\bibinfo
  {title} {Supersonic jet noise.}},\ }\href@noop {} {\bibfield  {journal}
  {\bibinfo  {journal} {Annu. Rev. Fluid Mech.}\ }\textbf {\bibinfo {volume}
  {27(1)}},\ \bibinfo {pages} {17} (\bibinfo {year} {1995})}\BibitemShut
  {NoStop}%
\bibitem [{\citenamefont {Miller}(2016)}]{Miller_2016}%
  \BibitemOpen
  \bibfield  {author} {\bibinfo {author} {\bibfnamefont {S.}~\bibnamefont
  {Miller}},\ }\bibfield  {title} {{\selectlanguage {English}\bibinfo {title}
  {Broadband shock-associated noise near-field cross-spectra}},\ }\href@noop {}
  {\bibfield  {journal} {\bibinfo  {journal} {J. Sound Vib.}\ }\textbf
  {\bibinfo {volume} {372}},\ \bibinfo {pages} {82} (\bibinfo {year}
  {2016})}\BibitemShut {NoStop}%
\bibitem [{\citenamefont {Norum}\ and\ \citenamefont
  {Seiner}(1982)}]{1982NormSeiner}%
  \BibitemOpen
  \bibfield  {author} {\bibinfo {author} {\bibfnamefont {T.~D.}\ \bibnamefont
  {Norum}}\ and\ \bibinfo {author} {\bibfnamefont {J.~M.}\ \bibnamefont
  {Seiner}},\ }\bibfield  {title} {{\selectlanguage {English}\bibinfo {title}
  {Broadband shock noise from supersonic jets}},\ }\href@noop {} {\bibfield
  {journal} {\bibinfo  {journal} {AIAA J.}\ }\textbf {\bibinfo {volume}
  {20(1)}},\ \bibinfo {pages} {68} (\bibinfo {year} {1982})}\BibitemShut
  {NoStop}%
\bibitem [{\citenamefont {Norum}(1984)}]{TDNorum_AIAA}%
  \BibitemOpen
  \bibfield  {author} {\bibinfo {author} {\bibfnamefont {T.~D.}\ \bibnamefont
  {Norum}},\ }\bibfield  {title} {\bibinfo {title} {{Control of jet shock
  associated noise by a reflector}},\ }in\ \href@noop {} {\emph {\bibinfo
  {booktitle} {9th Aeroacoustics Conference, AIAA Paper 1984-2279}}}\ (\bibinfo
  {year} {1984})\BibitemShut {NoStop}%
\bibitem [{\citenamefont {Ponton}\ \emph {et~al.}(1986)\citenamefont {Ponton},
  \citenamefont {Manning},\ and\ \citenamefont
  {Seiner}}]{farfield_rectangular}%
  \BibitemOpen
  \bibfield  {author} {\bibinfo {author} {\bibfnamefont {M.~K.}\ \bibnamefont
  {Ponton}}, \bibinfo {author} {\bibfnamefont {J.~C.}\ \bibnamefont
  {Manning}},\ and\ \bibinfo {author} {\bibfnamefont {J.~M.}\ \bibnamefont
  {Seiner}},\ }\bibfield  {title} {\bibinfo {title} {{Far-field acoustics of
  supersonic rectangular nozzles with various throat aspect ratios.}},\ }in\
  \href@noop {} {\emph {\bibinfo {booktitle} {NASA TM 89002}}}\ (\bibinfo
  {year} {1986})\BibitemShut {NoStop}%
\bibitem [{\citenamefont {Seiner}\ and\ \citenamefont
  {Ponton}(1986)}]{1986Seiner}%
  \BibitemOpen
  \bibfield  {author} {\bibinfo {author} {\bibfnamefont {J.~M.}\ \bibnamefont
  {Seiner}}\ and\ \bibinfo {author} {\bibfnamefont {M.~K.}\ \bibnamefont
  {Ponton}},\ }\bibfield  {title} {\bibinfo {title} {{Aeroacoustic Data for
  High Reynolds Number Supersonic Axisymmetric Jets.}},\ }in\ \href@noop {}
  {\emph {\bibinfo {booktitle} {NASA TM 86296}}}\ (\bibinfo {year}
  {1986})\BibitemShut {NoStop}%
\bibitem [{\citenamefont {Gutmark}\ \emph {et~al.}(1990)\citenamefont
  {Gutmark}, \citenamefont {Schadow},\ and\ \citenamefont
  {Bicker}}]{E.GUTMARK1990}%
  \BibitemOpen
  \bibfield  {author} {\bibinfo {author} {\bibfnamefont {E.}~\bibnamefont
  {Gutmark}}, \bibinfo {author} {\bibfnamefont {K.~C.}\ \bibnamefont
  {Schadow}},\ and\ \bibinfo {author} {\bibfnamefont {C.~J.}\ \bibnamefont
  {Bicker}},\ }\bibfield  {title} {{\selectlanguage {English}\bibinfo {title}
  {Near acoustic field and shock structure of rectangular supersonic jets.}},\
  }\href@noop {} {\bibfield  {journal} {\bibinfo  {journal} {AIAA J.}\ }\textbf
  {\bibinfo {volume} {28}},\ \bibinfo {pages} {1163} (\bibinfo {year}
  {1990})}\BibitemShut {NoStop}%
\bibitem [{\citenamefont {Wong}\ \emph {et~al.}(2019)\citenamefont {Wong},
  \citenamefont {Jordan},\ and\ \citenamefont {Honnery}}]{2019Wong}%
  \BibitemOpen
  \bibfield  {author} {\bibinfo {author} {\bibfnamefont {M.~H.}\ \bibnamefont
  {Wong}}, \bibinfo {author} {\bibfnamefont {P.}~\bibnamefont {Jordan}},\ and\
  \bibinfo {author} {\bibfnamefont {D.~R.}\ \bibnamefont {Honnery}},\
  }\bibfield  {title} {{\selectlanguage {English}\bibinfo {title} {Impact of
  coherence decay on wavepacket models for broadband shock-associated noise in
  supersonic jets}},\ }\href@noop {} {\bibfield  {journal} {\bibinfo  {journal}
  {J. Fluid Mech.}\ }\textbf {\bibinfo {volume} {863}},\ \bibinfo {pages} {969}
  (\bibinfo {year} {2019})}\BibitemShut {NoStop}%
\bibitem [{\citenamefont {Li}\ \emph {et~al.}(2019)\citenamefont {Li},
  \citenamefont {He}, \citenamefont {Zhang},\ and\ \citenamefont
  {Hao}}]{Xiangru_2019}%
  \BibitemOpen
  \bibfield  {author} {\bibinfo {author} {\bibfnamefont {X.}~\bibnamefont
  {Li}}, \bibinfo {author} {\bibfnamefont {F.}~\bibnamefont {He}}, \bibinfo
  {author} {\bibfnamefont {X.}~\bibnamefont {Zhang}},\ and\ \bibinfo {author}
  {\bibfnamefont {P.}~\bibnamefont {Hao}},\ }\bibfield  {title}
  {{\selectlanguage {English}\bibinfo {title} {Shock motion and flow structure
  of an underexpanded jet in the helical mode.}},\ }\href@noop {} {\bibfield
  {journal} {\bibinfo  {journal} {AIAA J.}\ }\textbf {\bibinfo {volume}
  {57(9)}},\ \bibinfo {pages} {3934} (\bibinfo {year} {2019})}\BibitemShut
  {NoStop}%
\bibitem [{\citenamefont {Nogueira}\ \emph
  {et~al.}(2022{\natexlab{a}})\citenamefont {Nogueira}, \citenamefont {Jordan},
  \citenamefont {Jaunet}, \citenamefont {Cavalieri}, \citenamefont {Towne},\
  and\ \citenamefont {Edgington-Mitchell}}]{absolute_instability}%
  \BibitemOpen
  \bibfield  {author} {\bibinfo {author} {\bibfnamefont {P.~A.~S.}\
  \bibnamefont {Nogueira}}, \bibinfo {author} {\bibfnamefont {P.}~\bibnamefont
  {Jordan}}, \bibinfo {author} {\bibfnamefont {V.}~\bibnamefont {Jaunet}},
  \bibinfo {author} {\bibfnamefont {A.~V.~G.}\ \bibnamefont {Cavalieri}},
  \bibinfo {author} {\bibfnamefont {A.}~\bibnamefont {Towne}},\ and\ \bibinfo
  {author} {\bibfnamefont {D.}~\bibnamefont {Edgington-Mitchell}},\ }\bibfield
  {title} {{\selectlanguage {English}\bibinfo {title} {Absolute instability in
  shock-containing jets}},\ }\href@noop {} {\bibfield  {journal} {\bibinfo
  {journal} {J. Fluid Mech.}\ }\textbf {\bibinfo {volume} {930}},\ \bibinfo
  {pages} {A10} (\bibinfo {year} {2022}{\natexlab{a}})}\BibitemShut {NoStop}%
\bibitem [{\citenamefont {Tam}\ and\ \citenamefont {Tanna}(1982)}]{1982Tam}%
  \BibitemOpen
  \bibfield  {author} {\bibinfo {author} {\bibfnamefont {C.~K.~W.}\
  \bibnamefont {Tam}}\ and\ \bibinfo {author} {\bibfnamefont {H.~K.}\
  \bibnamefont {Tanna}},\ }\bibfield  {title} {{\selectlanguage
  {English}\bibinfo {title} {Shock associated noise of supersonic jets from
  convergent-divergent nozzle}},\ }\href@noop {} {\bibfield  {journal}
  {\bibinfo  {journal} {J. Sound Vib.}\ }\textbf {\bibinfo {volume} {81(3)}},\
  \bibinfo {pages} {337} (\bibinfo {year} {1982})}\BibitemShut {NoStop}%
\bibitem [{\citenamefont {Pack}(1950)}]{1950Pack}%
  \BibitemOpen
  \bibfield  {author} {\bibinfo {author} {\bibfnamefont {D.~C.}\ \bibnamefont
  {Pack}},\ }\bibfield  {title} {{\selectlanguage {English}\bibinfo {title} {A
  note on $\rm p$randtl's formula for the wave-length of a supersonic gas
  jet}},\ }\href@noop {} {\bibfield  {journal} {\bibinfo  {journal}
  {\textsl{Qyart. Journ. Mech. and Applied Math.}}\ }\textbf {\bibinfo {volume}
  {3(2)}},\ \bibinfo {pages} {173} (\bibinfo {year} {1950})}\BibitemShut
  {NoStop}%
\bibitem [{\citenamefont {Gao}\ and\ \citenamefont {Li}(2010)}]{X.D.Li}%
  \BibitemOpen
  \bibfield  {author} {\bibinfo {author} {\bibfnamefont {J.~H.}\ \bibnamefont
  {Gao}}\ and\ \bibinfo {author} {\bibfnamefont {X.~D.}\ \bibnamefont {Li}},\
  }\bibfield  {title} {{\selectlanguage {English}\bibinfo {title} {A multi-mode
  screech frequency prediction formula for circular supersonic jets}},\
  }\href@noop {} {\bibfield  {journal} {\bibinfo  {journal} {J. Acoust. Soc.
  Am.}\ }\textbf {\bibinfo {volume} {127(3)}},\ \bibinfo {pages} {1251}
  (\bibinfo {year} {2010})}\BibitemShut {NoStop}%
\bibitem [{\citenamefont {Tam}(1988)}]{1988Tam_shockspacing}%
  \BibitemOpen
  \bibfield  {author} {\bibinfo {author} {\bibfnamefont {C.~K.~W.}\
  \bibnamefont {Tam}},\ }\bibfield  {title} {\bibinfo {title} {The shock-cell
  structures and screech tone frequencies of rectangular and non-axisymmetric
  supersonic jets.},\ }\href@noop {} {\bibfield  {journal} {\bibinfo  {journal}
  {J. Sound Vib.}\ }\textbf {\bibinfo {volume} {121(1)}},\ \bibinfo {pages}
  {135} (\bibinfo {year} {1988})}\BibitemShut {NoStop}%
\bibitem [{\citenamefont {Tam}\ and\ \citenamefont
  {Reddy}(1994)}]{1996Tam_JOA}%
  \BibitemOpen
  \bibfield  {author} {\bibinfo {author} {\bibfnamefont {C.~K.~W.}\
  \bibnamefont {Tam}}\ and\ \bibinfo {author} {\bibfnamefont {N.~N.}\
  \bibnamefont {Reddy}},\ }\bibfield  {title} {{\selectlanguage
  {English}\bibinfo {title} {Prediction method for broadband shock-associated
  noise from supersonic rectangular jets.}},\ }\href@noop {} {\bibfield
  {journal} {\bibinfo  {journal} {JOURNAL OF AIRCRAFT}\ }\textbf {\bibinfo
  {volume} {33(2)}},\ \bibinfo {pages} {298} (\bibinfo {year}
  {1994})}\BibitemShut {NoStop}%
\bibitem [{\citenamefont {Tam}\ \emph {et~al.}(1985)\citenamefont {Tam},
  \citenamefont {Jackson},\ and\ \citenamefont {Seiner}}]{1985Tam}%
  \BibitemOpen
  \bibfield  {author} {\bibinfo {author} {\bibfnamefont {C.~K.~W.}\
  \bibnamefont {Tam}}, \bibinfo {author} {\bibfnamefont {J.~A.}\ \bibnamefont
  {Jackson}},\ and\ \bibinfo {author} {\bibfnamefont {J.~M.}\ \bibnamefont
  {Seiner}},\ }\bibfield  {title} {{\selectlanguage {English}\bibinfo {title}
  {A multiple-scales model of the shock-cell structure of imperfectly expanded
  supersonic jets}},\ }\href@noop {} {\bibfield  {journal} {\bibinfo  {journal}
  {J. Fluid Mech.}\ }\textbf {\bibinfo {volume} {153}},\ \bibinfo {pages} {123}
  (\bibinfo {year} {1985})}\BibitemShut {NoStop}%
\bibitem [{\citenamefont {Song}\ \emph {et~al.}(2024)\citenamefont {Song},
  \citenamefont {Wu}, \citenamefont {Zhang},\ and\ \citenamefont
  {Fang}}]{2024Song_nonlinear}%
  \BibitemOpen
  \bibfield  {author} {\bibinfo {author} {\bibfnamefont {Z.}~\bibnamefont
  {Song}}, \bibinfo {author} {\bibfnamefont {X.}~\bibnamefont {Wu}}, \bibinfo
  {author} {\bibfnamefont {Z.}~\bibnamefont {Zhang}},\ and\ \bibinfo {author}
  {\bibfnamefont {Y.}~\bibnamefont {Fang}},\ }\bibfield  {title} {\bibinfo
  {title} {{The Weakly Nonlinear Development of Shock Cells in Screeching
  Jets.}},\ }in\ \href@noop {} {\emph {\bibinfo {booktitle} {30th AIAA
  Aeroacoustics Conference. AIAA Paper 24-3142}}}\ (\bibinfo {year}
  {2024})\BibitemShut {NoStop}%
\bibitem [{\citenamefont {Batchelor}\ and\ \citenamefont
  {Gill}(1962)}]{196batchelor}%
  \BibitemOpen
  \bibfield  {author} {\bibinfo {author} {\bibfnamefont {G.~K.}\ \bibnamefont
  {Batchelor}}\ and\ \bibinfo {author} {\bibfnamefont {A.~E.}\ \bibnamefont
  {Gill}},\ }\bibfield  {title} {{\selectlanguage {English}\bibinfo {title}
  {Analysis of the stability of axisymmetric jets}},\ }\href@noop {} {\bibfield
   {journal} {\bibinfo  {journal} {J. Fluid Mech.}\ }\textbf {\bibinfo {volume}
  {14(4)}},\ \bibinfo {pages} {529} (\bibinfo {year} {1962})}\BibitemShut
  {NoStop}%
\bibitem [{\citenamefont {Michalke}\ and\ \citenamefont
  {Hermann}(1982)}]{1982Alfons}%
  \BibitemOpen
  \bibfield  {author} {\bibinfo {author} {\bibfnamefont {A.}~\bibnamefont
  {Michalke}}\ and\ \bibinfo {author} {\bibfnamefont {G.}~\bibnamefont
  {Hermann}},\ }\bibfield  {title} {{\selectlanguage {English}\bibinfo {title}
  {On the inviscid instability of a circular jet with external flow}},\
  }\href@noop {} {\bibfield  {journal} {\bibinfo  {journal} {J. Fluid Mech.}\
  }\textbf {\bibinfo {volume} {114}},\ \bibinfo {pages} {343} (\bibinfo {year}
  {1982})}\BibitemShut {NoStop}%
\bibitem [{\citenamefont {Morris}(2010)}]{LSA_morris}%
  \BibitemOpen
  \bibfield  {author} {\bibinfo {author} {\bibfnamefont {P.~J.}\ \bibnamefont
  {Morris}},\ }\bibfield  {title} {{\selectlanguage {English}\bibinfo {title}
  {The instability of high speed jets}},\ }\href@noop {} {\bibfield  {journal}
  {\bibinfo  {journal} {International Journal of Aeroacoustics}\ }\textbf
  {\bibinfo {volume} {9(1-2)}},\ \bibinfo {pages} {1} (\bibinfo {year}
  {2010})}\BibitemShut {NoStop}%
\bibitem [{\citenamefont {Piot}\ \emph {et~al.}(2006)\citenamefont {Piot},
  \citenamefont {Casalis}, \citenamefont {Muller},\ and\ \citenamefont
  {Baily}}]{PSE_piot}%
  \BibitemOpen
  \bibfield  {author} {\bibinfo {author} {\bibfnamefont {E.}~\bibnamefont
  {Piot}}, \bibinfo {author} {\bibfnamefont {G.}~\bibnamefont {Casalis}},
  \bibinfo {author} {\bibfnamefont {F.}~\bibnamefont {Muller}},\ and\ \bibinfo
  {author} {\bibfnamefont {C.}~\bibnamefont {Baily}},\ }\bibfield  {title}
  {{\selectlanguage {English}\bibinfo {title} {Investigation of the pse
  approach for subsonic and supersonic hot jets. detailed comparisons with les
  and linearized euler equations results}},\ }\href@noop {} {\bibfield
  {journal} {\bibinfo  {journal} {Int. J. Aeroacoust.}\ }\textbf {\bibinfo
  {volume} {5(4)}},\ \bibinfo {pages} {361} (\bibinfo {year}
  {2006})}\BibitemShut {NoStop}%
\bibitem [{\citenamefont {Gudmundsson}\ and\ \citenamefont
  {Colonius}(2011)}]{Gudmundsson_jfm}%
  \BibitemOpen
  \bibfield  {author} {\bibinfo {author} {\bibfnamefont {K.}~\bibnamefont
  {Gudmundsson}}\ and\ \bibinfo {author} {\bibfnamefont {T.}~\bibnamefont
  {Colonius}},\ }\bibfield  {title} {{\selectlanguage {English}\bibinfo {title}
  {Instability wave models for the near-field fluctuations of turbulent
  jets}},\ }\href@noop {} {\bibfield  {journal} {\bibinfo  {journal} {J. Fluid
  Mech.}\ }\textbf {\bibinfo {volume} {689}},\ \bibinfo {pages} {97} (\bibinfo
  {year} {2011})}\BibitemShut {NoStop}%
\bibitem [{\citenamefont {Nogueira}\ \emph
  {et~al.}(2022{\natexlab{b}})\citenamefont {Nogueira}, \citenamefont {Self},
  \citenamefont {Towne},\ and\ \citenamefont
  {Edgington-Mitchell}}]{nogueira2022wave}%
  \BibitemOpen
  \bibfield  {author} {\bibinfo {author} {\bibfnamefont {P.~A.}\ \bibnamefont
  {Nogueira}}, \bibinfo {author} {\bibfnamefont {H.~W.}\ \bibnamefont {Self}},
  \bibinfo {author} {\bibfnamefont {A.}~\bibnamefont {Towne}},\ and\ \bibinfo
  {author} {\bibfnamefont {D.}~\bibnamefont {Edgington-Mitchell}},\ }\bibfield
  {title} {\bibinfo {title} {Wave-packet modulation in shock-containing jets},\
  }\href@noop {} {\bibfield  {journal} {\bibinfo  {journal} {Phys. Rev. Fluid}\
  }\textbf {\bibinfo {volume} {7(7)}},\ \bibinfo {pages} {074608} (\bibinfo
  {year} {2022}{\natexlab{b}})}\BibitemShut {NoStop}%
\bibitem [{\citenamefont {Crighton}\ and\ \citenamefont
  {Gaster}(1976)}]{1976Crighton_diverging}%
  \BibitemOpen
  \bibfield  {author} {\bibinfo {author} {\bibfnamefont {D.~G.}\ \bibnamefont
  {Crighton}}\ and\ \bibinfo {author} {\bibfnamefont {M.}~\bibnamefont
  {Gaster}},\ }\bibfield  {title} {{\selectlanguage {English}\bibinfo {title}
  {Stability of slowly diverging jet flow}},\ }\href@noop {} {\bibfield
  {journal} {\bibinfo  {journal} {J. Fluid Mech.}\ }\textbf {\bibinfo {volume}
  {77}},\ \bibinfo {pages} {397} (\bibinfo {year} {1976})}\BibitemShut
  {NoStop}%
\bibitem [{\citenamefont {Yen}\ and\ \citenamefont
  {Messersmith}(1998)}]{Yen_aiaaj}%
  \BibitemOpen
  \bibfield  {author} {\bibinfo {author} {\bibfnamefont {C.~C.}\ \bibnamefont
  {Yen}}\ and\ \bibinfo {author} {\bibfnamefont {N.~L.}\ \bibnamefont
  {Messersmith}},\ }\bibfield  {title} {{\selectlanguage {English}\bibinfo
  {title} {Application of parabolized stability equations to the prediction of
  jet instabilities}},\ }\href@noop {} {\bibfield  {journal} {\bibinfo
  {journal} {AIAA J.}\ }\textbf {\bibinfo {volume} {36(8)}},\ \bibinfo {pages}
  {1541} (\bibinfo {year} {1998})}\BibitemShut {NoStop}%
\bibitem [{\citenamefont {Towne}\ \emph {et~al.}(2022)\citenamefont {Towne},
  \citenamefont {Rigas}, \citenamefont {Kamal}, \citenamefont {Pickering},\
  and\ \citenamefont {Colonius}}]{OWNSE_2022}%
  \BibitemOpen
  \bibfield  {author} {\bibinfo {author} {\bibfnamefont {A.}~\bibnamefont
  {Towne}}, \bibinfo {author} {\bibfnamefont {G.}~\bibnamefont {Rigas}},
  \bibinfo {author} {\bibfnamefont {O.}~\bibnamefont {Kamal}}, \bibinfo
  {author} {\bibfnamefont {E.}~\bibnamefont {Pickering}},\ and\ \bibinfo
  {author} {\bibfnamefont {T.}~\bibnamefont {Colonius}},\ }\bibfield  {title}
  {{\selectlanguage {English}\bibinfo {title} {Efficient global resolvent
  analysis via the one-way navier–stokes equations}},\ }\href@noop {}
  {\bibfield  {journal} {\bibinfo  {journal} {Journal of Fluid Mechanics}\
  }\textbf {\bibinfo {volume} {948}},\ \bibinfo {pages} {A9} (\bibinfo {year}
  {2022})}\BibitemShut {NoStop}%
\bibitem [{\citenamefont {Wu}(2005)}]{Wu_nonlinearMachWave}%
  \BibitemOpen
  \bibfield  {author} {\bibinfo {author} {\bibfnamefont {X.}~\bibnamefont
  {Wu}},\ }\bibfield  {title} {{\selectlanguage {English}\bibinfo {title} {Mach
  wave radiation of nonlinearly evolving supersonic instability modes in shear
  layers}},\ }\href@noop {} {\bibfield  {journal} {\bibinfo  {journal} {J.
  Fluid Mech.}\ }\textbf {\bibinfo {volume} {523}},\ \bibinfo {pages} {121}
  (\bibinfo {year} {2005})}\BibitemShut {NoStop}%
\bibitem [{\citenamefont {Wu}(2019)}]{Wu_annual}%
  \BibitemOpen
  \bibfield  {author} {\bibinfo {author} {\bibfnamefont {X.}~\bibnamefont
  {Wu}},\ }\bibfield  {title} {{\selectlanguage {English}\bibinfo {title}
  {Nonlinear theories for shear flow instabilities: Physical insights and
  practical implications}},\ }\href@noop {} {\bibfield  {journal} {\bibinfo
  {journal} {Annu. Rev. Fluid Mech.}\ }\textbf {\bibinfo {volume} {51}},\
  \bibinfo {pages} {451} (\bibinfo {year} {2019})}\BibitemShut {NoStop}%
\bibitem [{\citenamefont {Jordan}\ and\ \citenamefont
  {Colonius}(2013)}]{2013_annual_rev}%
  \BibitemOpen
  \bibfield  {author} {\bibinfo {author} {\bibfnamefont {P.}~\bibnamefont
  {Jordan}}\ and\ \bibinfo {author} {\bibfnamefont {T.}~\bibnamefont
  {Colonius}},\ }\bibfield  {title} {{\selectlanguage {English}\bibinfo {title}
  {Wave packets and turbulent jet noise.}},\ }\href@noop {} {\bibfield
  {journal} {\bibinfo  {journal} {Annu. Rev. Fluid Mech.}\ }\textbf {\bibinfo
  {volume} {45(1)}},\ \bibinfo {pages} {173} (\bibinfo {year}
  {2013})}\BibitemShut {NoStop}%
\bibitem [{\citenamefont {Pickering}\ \emph {et~al.}(2020)\citenamefont
  {Pickering}, \citenamefont {Rigas}, \citenamefont {Nogueira}, \citenamefont
  {Cavalieri}, \citenamefont {Schmidt},\ and\ \citenamefont
  {Colonius}}]{2020_jfm_wavepackets}%
  \BibitemOpen
  \bibfield  {author} {\bibinfo {author} {\bibfnamefont {E.}~\bibnamefont
  {Pickering}}, \bibinfo {author} {\bibfnamefont {G.}~\bibnamefont {Rigas}},
  \bibinfo {author} {\bibfnamefont {P.~A.~S.}\ \bibnamefont {Nogueira}},
  \bibinfo {author} {\bibfnamefont {A.~V.~G.}\ \bibnamefont {Cavalieri}},
  \bibinfo {author} {\bibfnamefont {O.~T.}\ \bibnamefont {Schmidt}},\ and\
  \bibinfo {author} {\bibfnamefont {T.}~\bibnamefont {Colonius}},\ }\bibfield
  {title} {{\selectlanguage {English}\bibinfo {title} {Ilift-up,
  kelvin-helmholtz and orr mechanisms in turbulent jets}},\ }\href@noop {}
  {\bibfield  {journal} {\bibinfo  {journal} {J. Fluid Mech.}\ }\textbf
  {\bibinfo {volume} {896}},\ \bibinfo {pages} {A2} (\bibinfo {year}
  {2020})}\BibitemShut {NoStop}%
\bibitem [{\citenamefont {Maia}\ \emph {et~al.}(2019)\citenamefont {Maia},
  \citenamefont {Jordan}, \citenamefont {Cavalieri},\ and\ \citenamefont
  {Jaunet}}]{exponetially_Gaussian}%
  \BibitemOpen
  \bibfield  {author} {\bibinfo {author} {\bibfnamefont {I.~A.}\ \bibnamefont
  {Maia}}, \bibinfo {author} {\bibfnamefont {P.}~\bibnamefont {Jordan}},
  \bibinfo {author} {\bibfnamefont {A.~V.~G.}\ \bibnamefont {Cavalieri}},\ and\
  \bibinfo {author} {\bibfnamefont {V.}~\bibnamefont {Jaunet}},\ }\bibfield
  {title} {{\selectlanguage {English}\bibinfo {title} {Two-point wavepacket
  modelling of jet noise}},\ }\href@noop {} {\bibfield  {journal} {\bibinfo
  {journal} {Proc. R. Soc. A}\ }\textbf {\bibinfo {volume} {475(2227)}},\
  \bibinfo {pages} {20190199} (\bibinfo {year} {2019})}\BibitemShut {NoStop}%
\bibitem [{\citenamefont {Harper-Bourne}\ and\ \citenamefont
  {Fisher}(1973)}]{1973Harper}%
  \BibitemOpen
  \bibfield  {author} {\bibinfo {author} {\bibfnamefont {M.}~\bibnamefont
  {Harper-Bourne}}\ and\ \bibinfo {author} {\bibfnamefont {M.~J.}\ \bibnamefont
  {Fisher}},\ }\bibfield  {title} {{\selectlanguage {English}\bibinfo {title}
  {The noise from shock waves in supersonic jets.}},\ }\href@noop {} {\bibfield
   {journal} {\bibinfo  {journal} {AGARD Technical Report CP-131}\ }\textbf
  {\bibinfo {volume} {11}},\ \bibinfo {pages} {1} (\bibinfo {year}
  {1973})}\BibitemShut {NoStop}%
\bibitem [{\citenamefont {Tam}(1987)}]{1988Tam}%
  \BibitemOpen
  \bibfield  {author} {\bibinfo {author} {\bibfnamefont {C.~K.~W.}\
  \bibnamefont {Tam}},\ }\bibfield  {title} {{\selectlanguage {English}\bibinfo
  {title} {Stochastic model theory of broadband shock associated noise from
  supersonic jets}},\ }\href@noop {} {\bibfield  {journal} {\bibinfo  {journal}
  {\textsl{J. Sound Vib.}}\ }\textbf {\bibinfo {volume} {116(2)}},\ \bibinfo
  {pages} {265} (\bibinfo {year} {1987})}\BibitemShut {NoStop}%
\bibitem [{\citenamefont {Lele}(2005)}]{2005Lele}%
  \BibitemOpen
  \bibfield  {author} {\bibinfo {author} {\bibfnamefont {S.~K.}\ \bibnamefont
  {Lele}},\ }\bibfield  {title} {\bibinfo {title} {{Phased array models of
  shock-cell noise sources.}},\ }in\ \href@noop {} {\emph {\bibinfo {booktitle}
  {11th AIAA/CEAS Aeroacoustics Conference. AIAA Paper 2005-2841}}}\ (\bibinfo
  {year} {2005})\BibitemShut {NoStop}%
\bibitem [{\citenamefont {Lighthill}(1952)}]{1952lighthill}%
  \BibitemOpen
  \bibfield  {author} {\bibinfo {author} {\bibfnamefont {M.~J.}\ \bibnamefont
  {Lighthill}},\ }\bibfield  {title} {{\selectlanguage {English}\bibinfo
  {title} {On sound generated aerodynamically.
  \uppercase\expandafter{\romannumeral1}. general theory}},\ }\href@noop {}
  {\bibfield  {journal} {\bibinfo  {journal} {Proc. R. Soc. Lond. A}\ }\textbf
  {\bibinfo {volume} {211(1107)}},\ \bibinfo {pages} {564} (\bibinfo {year}
  {1952})}\BibitemShut {NoStop}%
\bibitem [{\citenamefont {Tam}\ \emph {et~al.}(1986)\citenamefont {Tam},
  \citenamefont {Seiner},\ and\ \citenamefont {YU}}]{1986Tam_Proposed}%
  \BibitemOpen
  \bibfield  {author} {\bibinfo {author} {\bibfnamefont {C.~K.~W.}\
  \bibnamefont {Tam}}, \bibinfo {author} {\bibfnamefont {J.~M.}\ \bibnamefont
  {Seiner}},\ and\ \bibinfo {author} {\bibfnamefont {J.~C.}\ \bibnamefont
  {YU}},\ }\bibfield  {title} {\bibinfo {title} {Proposed relationship between
  broadband shock associated noise and screech tones.},\ }\href@noop {}
  {\bibfield  {journal} {\bibinfo  {journal} {J. Sound Vib.}\ }\textbf
  {\bibinfo {volume} {110(2)}},\ \bibinfo {pages} {309} (\bibinfo {year}
  {1986})}\BibitemShut {NoStop}%
\bibitem [{\citenamefont {Cavalieri}\ \emph {et~al.}(2019)\citenamefont
  {Cavalieri}, \citenamefont {Jordan},\ and\ \citenamefont
  {Lesshafft}}]{2019wavepacket}%
  \BibitemOpen
  \bibfield  {author} {\bibinfo {author} {\bibfnamefont {A.~V.~G.}\
  \bibnamefont {Cavalieri}}, \bibinfo {author} {\bibfnamefont {P.}~\bibnamefont
  {Jordan}},\ and\ \bibinfo {author} {\bibfnamefont {L.}~\bibnamefont
  {Lesshafft}},\ }\bibfield  {title} {{\selectlanguage {English}\bibinfo
  {title} {Wave-packet models for jet dynamics and sound radiation}},\
  }\href@noop {} {\bibfield  {journal} {\bibinfo  {journal} {Applied Mechanics
  Reviews}\ }\textbf {\bibinfo {volume} {71(2)}},\ \bibinfo {pages} {020802}
  (\bibinfo {year} {2019})}\BibitemShut {NoStop}%
\bibitem [{\citenamefont {Li}\ and\ \citenamefont {Lyu}(2023)}]{MyOwn_2}%
  \BibitemOpen
  \bibfield  {author} {\bibinfo {author} {\bibfnamefont {B.}~\bibnamefont
  {Li}}\ and\ \bibinfo {author} {\bibfnamefont {B.}~\bibnamefont {Lyu}},\
  }\bibfield  {title} {{\selectlanguage {English}\bibinfo {title} {Acoustic
  emission due to the interaction between shock and instability waves in
  two-dimensional supersonic jet flows}},\ }\href@noop {} {\bibfield  {journal}
  {\bibinfo  {journal} {J. Fluid Mech.}\ }\textbf {\bibinfo {volume} {954}},\
  \bibinfo {pages} {A35} (\bibinfo {year} {2023})}\BibitemShut {NoStop}%
\bibitem [{\citenamefont {Noble}(1958)}]{Noble}%
  \BibitemOpen
  \bibfield  {author} {\bibinfo {author} {\bibfnamefont {B.}~\bibnamefont
  {Noble}},\ }\bibinfo {title} {Methos based on the wiener-hopf technique}\
  (\bibinfo  {publisher} {Nover},\ \bibinfo {address} {New York},\ \bibinfo
  {year} {1958})\ \bibinfo {edition} {3rd}\ ed.\BibitemShut {Stop}%
\bibitem [{\citenamefont {Crighton}\ \emph {et~al.}(1992)\citenamefont
  {Crighton}, \citenamefont {Dowling}, \citenamefont {Williams}, \citenamefont
  {Heckl},\ and\ \citenamefont {Leppington}}]{ModernMethods}%
  \BibitemOpen
  \bibfield  {author} {\bibinfo {author} {\bibfnamefont {D.~G.}\ \bibnamefont
  {Crighton}}, \bibinfo {author} {\bibfnamefont {A.~P.}\ \bibnamefont
  {Dowling}}, \bibinfo {author} {\bibfnamefont {J.~F.}\ \bibnamefont
  {Williams}}, \bibinfo {author} {\bibfnamefont {M.~A.}\ \bibnamefont
  {Heckl}},\ and\ \bibinfo {author} {\bibfnamefont {F.~A.}\ \bibnamefont
  {Leppington}},\ }\bibinfo {title} {Modern methods in analytical acoustics:
  Lecture notes}\ (\bibinfo  {publisher} {Springer-Verlag},\ \bibinfo {address}
  {London},\ \bibinfo {year} {1992})\ Chap.\ \bibinfo {chapter} {4.4},\
  \bibinfo {edition} {1st}\ ed.\BibitemShut {Stop}%
\bibitem [{\citenamefont {Zaman}(1998)}]{zaman1998asymptotic}%
  \BibitemOpen
  \bibfield  {author} {\bibinfo {author} {\bibfnamefont {K.}~\bibnamefont
  {Zaman}},\ }\bibfield  {title} {\bibinfo {title} {Asymptotic spreading rate
  of initially compressible jets—experiment and analysis},\ }\href@noop {}
  {\bibfield  {journal} {\bibinfo  {journal} {Physics of Fluids}\ }\textbf
  {\bibinfo {volume} {10(10)}},\ \bibinfo {pages} {2652} (\bibinfo {year}
  {1998})}\BibitemShut {NoStop}%
\bibitem [{\citenamefont {Panda}\ and\ \citenamefont
  {Seasholtz}(1999)}]{shock_spacing_panda}%
  \BibitemOpen
  \bibfield  {author} {\bibinfo {author} {\bibfnamefont {J.}~\bibnamefont
  {Panda}}\ and\ \bibinfo {author} {\bibfnamefont {R.~G.}\ \bibnamefont
  {Seasholtz}},\ }\bibfield  {title} {{\selectlanguage {English}\bibinfo
  {title} {Measurement of shock structure and shock–vortex interaction in
  underexpanded jets using rayleigh scattering.}},\ }\href@noop {} {\bibfield
  {journal} {\bibinfo  {journal} {Phys. Fluids}\ }\textbf {\bibinfo {volume}
  {11}},\ \bibinfo {pages} {3761} (\bibinfo {year} {1999})}\BibitemShut
  {NoStop}%
\bibitem [{\citenamefont {Edgington-Mitchell}\ \emph
  {et~al.}(2014)\citenamefont {Edgington-Mitchell}, \citenamefont
  {Oberleithner}, \citenamefont {Honnery},\ and\ \citenamefont
  {Soria}}]{2014edgington}%
  \BibitemOpen
  \bibfield  {author} {\bibinfo {author} {\bibfnamefont {D.}~\bibnamefont
  {Edgington-Mitchell}}, \bibinfo {author} {\bibfnamefont {K.}~\bibnamefont
  {Oberleithner}}, \bibinfo {author} {\bibfnamefont {D.~R.}\ \bibnamefont
  {Honnery}},\ and\ \bibinfo {author} {\bibfnamefont {J.}~\bibnamefont
  {Soria}},\ }\bibfield  {title} {{\selectlanguage {English}\bibinfo {title}
  {Coherent structure and sound production in the helical mode of a screeching
  axisymmetric jet}},\ }\href@noop {} {\bibfield  {journal} {\bibinfo
  {journal} {J. Fluid Mech.}\ }\textbf {\bibinfo {volume} {748}},\ \bibinfo
  {pages} {822} (\bibinfo {year} {2014})}\BibitemShut {NoStop}%
\bibitem [{\citenamefont {Edgington-Mitchell}\ \emph
  {et~al.}(2021)\citenamefont {Edgington-Mitchell}, \citenamefont {Weightman},
  \citenamefont {Lock}, \citenamefont {Kirby}, \citenamefont {Nair},
  \citenamefont {Soria}, ,\ and\ \citenamefont
  {Honnery}}]{edgington_shockleakage}%
  \BibitemOpen
  \bibfield  {author} {\bibinfo {author} {\bibfnamefont {D.}~\bibnamefont
  {Edgington-Mitchell}}, \bibinfo {author} {\bibfnamefont {J.}~\bibnamefont
  {Weightman}}, \bibinfo {author} {\bibfnamefont {S.}~\bibnamefont {Lock}},
  \bibinfo {author} {\bibfnamefont {R.}~\bibnamefont {Kirby}}, \bibinfo
  {author} {\bibfnamefont {V.}~\bibnamefont {Nair}}, \bibinfo {author}
  {\bibfnamefont {J.}~\bibnamefont {Soria}}, ,\ and\ \bibinfo {author}
  {\bibfnamefont {D.}~\bibnamefont {Honnery}},\ }\bibfield  {title}
  {{\selectlanguage {English}\bibinfo {title} {The generation of screech tones
  by shock leakage.}},\ }\href@noop {} {\bibfield  {journal} {\bibinfo
  {journal} {J. Fluid Mech.}\ }\textbf {\bibinfo {volume} {908}},\ \bibinfo
  {pages} {A46} (\bibinfo {year} {2021})}\BibitemShut {NoStop}%
\bibitem [{\citenamefont {Semlitsch}\ \emph {et~al.}(2020)\citenamefont
  {Semlitsch}, \citenamefont {Malla}, \citenamefont {Gutmark},\ and\
  \citenamefont {Mihaescu}}]{2020_sound_sources}%
  \BibitemOpen
  \bibfield  {author} {\bibinfo {author} {\bibfnamefont {B.}~\bibnamefont
  {Semlitsch}}, \bibinfo {author} {\bibfnamefont {B.}~\bibnamefont {Malla}},
  \bibinfo {author} {\bibfnamefont {E.~J.}\ \bibnamefont {Gutmark}},\ and\
  \bibinfo {author} {\bibfnamefont {M.}~\bibnamefont {Mihaescu}},\ }\bibfield
  {title} {{\selectlanguage {English}\bibinfo {title} {The generation mechanism
  of higher screech tone harmonics in supersonic jets.}},\ }\href@noop {}
  {\bibfield  {journal} {\bibinfo  {journal} {J. Fluid Mech.}\ }\textbf
  {\bibinfo {volume} {893}},\ \bibinfo {pages} {A9} (\bibinfo {year}
  {2020})}\BibitemShut {NoStop}%
\bibitem [{\citenamefont {Yu}\ and\ \citenamefont {Dosanjh}(1972)}]{Yu_1972}%
  \BibitemOpen
  \bibfield  {author} {\bibinfo {author} {\bibfnamefont {J.~C.}\ \bibnamefont
  {Yu}}\ and\ \bibinfo {author} {\bibfnamefont {D.~S.}\ \bibnamefont
  {Dosanjh}},\ }\bibfield  {title} {{\selectlanguage {English}\bibinfo {title}
  {Noise field of a supersonic mach 1.5 cold model jet}},\ }\href@noop {}
  {\bibfield  {journal} {\bibinfo  {journal} {The Journal of the Acoustical
  Society of America}\ }\textbf {\bibinfo {volume} {51(5A)}},\ \bibinfo {pages}
  {1400} (\bibinfo {year} {1972})}\BibitemShut {NoStop}%
\bibitem [{\citenamefont {Edgington-Mitchell}(2019)}]{2019EDINGTON}%
  \BibitemOpen
  \bibfield  {author} {\bibinfo {author} {\bibfnamefont {D.}~\bibnamefont
  {Edgington-Mitchell}},\ }\bibfield  {title} {{\selectlanguage
  {English}\bibinfo {title} {Aeroacoustic resonance and self-excitation in
  screeching and impinging supersonic jets – a review}},\ }\href@noop {}
  {\bibfield  {journal} {\bibinfo  {journal} {International Journal of
  Aeroacoustics}\ }\textbf {\bibinfo {volume} {18(2-3)}},\ \bibinfo {pages}
  {118} (\bibinfo {year} {2019})}\BibitemShut {NoStop}%
\bibitem [{\citenamefont {Lyu}\ and\ \citenamefont {Dowling}(2017)}]{lyu2017}%
  \BibitemOpen
  \bibfield  {author} {\bibinfo {author} {\bibfnamefont {B.}~\bibnamefont
  {Lyu}}\ and\ \bibinfo {author} {\bibfnamefont {A.~P.}\ \bibnamefont
  {Dowling}},\ }\bibfield  {title} {\bibinfo {title} {{On the Mechanism and
  Reduction of Installed Jet Noise}},\ }in\ \href@noop {} {\emph {\bibinfo
  {booktitle} {23rd AIAA/CEAS Aeroacoustics Conference, AIAA Paper
  2017-3523}}}\ (\bibinfo {year} {2017})\BibitemShut {NoStop}%
\bibitem [{\citenamefont {Lyu}\ and\ \citenamefont {Dowling}(2018)}]{lyu2018}%
  \BibitemOpen
  \bibfield  {author} {\bibinfo {author} {\bibfnamefont {B.}~\bibnamefont
  {Lyu}}\ and\ \bibinfo {author} {\bibfnamefont {A.~P.}\ \bibnamefont
  {Dowling}},\ }\bibfield  {title} {\bibinfo {title} {{Prediction of installed
  jet noise due to swept wings}},\ }in\ \href@noop {} {\emph {\bibinfo
  {booktitle} {24th AIAA/CEAS Aeroacoustics Conference, AIAA Paper
  2018-2980}}}\ (\bibinfo {year} {2018})\BibitemShut {NoStop}%
\bibitem [{\citenamefont {Lyu}\ and\ \citenamefont
  {Dowling}(2019)}]{Lyu_Dowling_2019}%
  \BibitemOpen
  \bibfield  {author} {\bibinfo {author} {\bibfnamefont {B.}~\bibnamefont
  {Lyu}}\ and\ \bibinfo {author} {\bibfnamefont {A.~P.}\ \bibnamefont
  {Dowling}},\ }\bibfield  {title} {\bibinfo {title} {Modelling installed jet
  noise due to the scattering of jet instability waves by swept wings},\
  }\href@noop {} {\bibfield  {journal} {\bibinfo  {journal} {Journal of Fluid
  Mechanics}\ }\textbf {\bibinfo {volume} {870}},\ \bibinfo {pages} {760–783}
  (\bibinfo {year} {2019})}\BibitemShut {NoStop}%
\bibitem [{\citenamefont {Tam}\ and\ \citenamefont
  {Burton}(1984)}]{Tam_Burton_1984}%
  \BibitemOpen
  \bibfield  {author} {\bibinfo {author} {\bibfnamefont {C.~K.~W.}\
  \bibnamefont {Tam}}\ and\ \bibinfo {author} {\bibfnamefont {D.~E.}\
  \bibnamefont {Burton}},\ }\bibfield  {title} {\bibinfo {title} {Sound
  generated by instability waves of supersonic flows. part 2. axisymmetric
  jets},\ }\href@noop {} {\bibfield  {journal} {\bibinfo  {journal} {Journal of
  Fluid Mechanics}\ }\textbf {\bibinfo {volume} {138}},\ \bibinfo {pages}
  {273–295} (\bibinfo {year} {1984})}\BibitemShut {NoStop}%
\end{thebibliography}%

\end{document}